\documentclass[a4paper,11pt]{article}
\pdfoutput=1 % if your are submitting a pdflatex (i.e. if you have
\usepackage{jheppub} % for details on the use of the package, please
\usepackage[T1]{fontenc}
\usepackage[utf8]{inputenc}

\usepackage{graphicx, wrapfig, dcolumn}
\usepackage{color, subfigure, slashed, listings, fancyhdr, amsmath, amsthm, amssymb, cmap}

\title{Running coupling in the conformal window of large-Nf QCD}

\author[a,1]{Markus Hopfer,\note{Corresponding author.}}
\author[b]{Christian S. Fischer,}
\author[a]{Reinhard Alkofer}

\affiliation[a]{Institut f\"ur Physik, Universit\"at Graz,\\Universit\"atsplatz 5, 8010 Graz, Austria}
\affiliation[b]{Institut f\"ur Theoretische Physik, Justus-Liebig Universit\"at Giessen,\\Heinrich-Buff-Ring 16, 35392 Giessen, Germany}

\emailAdd{markus.hopfer@uni-graz.at}
\emailAdd{Christian.Fischer@theo.physik.uni-giessen.de}
\emailAdd{reinhard.alkofer@uni-graz.at}

\abstract{Quantum Chromodynamics
with a relatively large number of fundamentally charged quark flavours
in the chiral limit is considered. A self-consistent solution of the
quark, gluon and ghost propagator Dyson-Schwinger equations in Landau gauge
exhibits a phase transition.  Above the critical number of fermion flavours
the non-perturbative running coupling develops a plateau over a wide momentum
range, and the propagators follow a power law behaviour for these momenta.
Hereby, the critical number of quark flavours depends crucially on the
beyond-tree-level tensor structures of the quark-gluon vertex.
}

\begin{document}
\maketitle
\flushbottom

% \PACS{11.15.-q, 11.30.Rd, 12.38.Aw}
%%%%%%%%%%%%%%%%%%%%%%%%%%%%%%%%%%%%%%%%%%%%%%%%%%%%%%%%%%%%%%%%%%%%%%%%%%%%%%%%%%%%%%%%%%%%%%%%%%%%%%%%%%%%%%%%%  
\section{Introduction}
Extensions of Quantum Chromodynamics (QCD), and hereby especially the
infrared (IR) behaviour of related gauge theories, are studied for a number of 
reasons.
The here presented study of gauge boson and fermion propagators (dubbed for 
simplicity gluon and quark propagators in the following) for a relatively large 
number of fundamentally charged quark flavours in the chiral limit is 
motivated mainly by two issues.
First, gauge theories with a near-conformal behaviour 
have attracted quite some attention as a possibility to break the electroweak 
symmetry dynamically without an obvious conflict to observation (which so far is in 
perfect agreement with the Standard Model), see {\it e.g.} \cite{Farhi:1980xs} and references therein. 
These so-called walking technicolor models provide a dynamical 
(vs.\ spontaneous) Higgs mechanism that accounts for the Higgs scalar as a bound state
of fermions and generates the mass for the electroweak gauge bosons ($W$ and $Z$) 
through a new hypothetical gauge interaction (technicolor) coupled to new 
massless fermions (techniquarks). Their chiral symmetry allows to avoid
fine-tuning of the Higgs mass and consequently the related hierarchy problem 
would be resolved.
Second, understanding the phase diagram of QCD for non-vanishing external 
parameters, like temperature and chemical potential, has proven to be a very hard
problem. Therefore phase transitions occurring as a function of group and 
representation properties are of special interest as they may help understanding
aspects of the QCD phase diagram in a technically simpler setting. 

From the point of view of Beyond-Standard-Model (BSM) physics
one should note that the hypothesized new gauge interaction is typically
required to behave  asymptotically free at scales much higher than the
electroweak one, then to  become near-conformal in an intermediate range, and 
strong and confining at lower energies such that the chiral symmetry of the
massless fermions is dynamically broken. This motivates the interest in the 
lower region of the so-called conformal window as well as its bordering 
confining counterpart. The above assumptions are made to overcome serious 
phenomenological problems inherent to early technicolor models,
as originally proposed in \cite{Weinberg:1979bn} or extended in 
\cite{Dimopoulos:1979es,Eichten:1979ah}. 
Contrary to the {Standard Model} interactions the newly introduced gauge 
interactions impose in principle no restrictions on fermion flavour mixing. 
The {Standard Model} flavour changing neutral currents (FCNCs) are not
only absent at tree-level and thus suppressed but this  suppression is 
also amplified by the Glashow-Iliopoulos-Maiani mechanism 
\cite{Glashow:1970gm}.  
Experimental measurements of FCNC processes via, {\it e.g.}, 
rare B meson decays confirm this Standard Model prediction 
and thus put stringent bounds on technicolor scenarios.

Walking technicolor models may overcome at least some of these obstacles, 
{\it see e.g.},  
\cite{Holdom:1984sk,Yamawaki:1985zg,Appelquist:1986an,Lane:2002wv,Sannino:2009za}
 and references therein. 
These models exhibit an approximate scale invariance over a wide energy range 
as well as a proximity (in parameter space) to an infrared fixed point 
(IRFP). Correspondingly, the gauge coupling is slowly running, or ``walking'' 
\cite{Appelquist:1986an,Lane:1991qh,Appelquist:1997fp}.
The relation to QCD occurs through the observation that asymptotically 
free SU$(N_c)$ gauge theories with a relatively large number of 
massless fermions can possess such properties \cite{Lane:1991qh}.
In QCD the self-interaction between the gluons and the related quantum fluctuations 
provide anti-screening while quarks are screening in the same way as electrons are
in QED. As long as the number of light flavours in QCD is small the gauge
coupling increases with decreasing scale and (as evident from hadron phenomenology 
and substantiated by lattice calculations) the  interactions between quarks
are strong enough in the intermediate and IR momentum regime
to trigger dynamical chiral symmetry breaking as well as confinement. 
On the other hand, just below the maximal flavour number 
$N_f^{AF}=11N_c / 2 = 16.5$, 
at which QCD loses its asymptotic freedom, the theory behaves conformal and 
develops an IRFP \cite{Caswell:1974,Banks:1982}.
When lowering the number of light, resp.\ massless, flavours, $N_f$, the strength 
of this IRFP increases further. However, at some critical number of flavours,
$N_f^{crit}$, the transition to the chirally broken and confining phase must occur.
The value $N_f^{crit}$ then defines the above mentioned
lower bound of the conformal window.
By now it seems generally accepted that $N_f=12$ lies inside the conformal window 
\cite{Appelquist:1996dq,Gies:2005as,Appelquist:2007hu,Appelquist:2009ty} (see, however, 
ref.~\cite{Fodor:2012et}) but a precise value for the lower bound of the conformal 
window of QCD is still unknown.

Based on equations for the Green functions of a given quantum field theory 
functional methods provide appropriate non-perturbative tools to explore the 
theory over a wide momentum range. For asymptotically free gauge theories
they are able to connect the deep IR with the perturbative ultraviolet (UV) 
regime. Within the framework of the functional
renormalization group exact scaling relations for physical observables at or
close to the quantum critical point at $N_f^{crit}$ have been identified in 
ref.~\cite{Braun:2009ns,Braun:2010qs} and used to derive the leading order
scaling behaviour of infrared observables at $N_f^{crit}$. Furthermore, based
on an elaborate truncation scheme for the effective action an estimate of the 
critical number of fermion flavours of $N_f^{crit} \approx 10$ has been given.

In this work we will employ the Dyson-Schwinger framework, see {\it e.g.}, 
Refs.~\cite{Alkofer:2000wg,Maris:2003vk} and references therein.
Since Dyson-Schwinger equations (DSEs) constitute an infinite set of 
coupled integral equations carefully chosen truncations have to be applied 
in order to obtain sensible results for all (Euclidean) momenta. 
In this context it is 
useful that in the limit of small Euclidean momenta  it is even possible to 
solve the whole tower of equations self-consistently without relying on any kind 
of truncation \cite{Alkofer:2004it}. 
This allows one to put constraints onto the IR behaviour of the theory. 
In the UV regime, on the other hand, asymptotic freedom guarantees a 
unique determination of the Green functions given by their perturbative 
anomalous dimensions.
Nevertheless, a comparison with other non-perturbative methods as, {\it e.g.}, 
lattice is inevitable at some point in order to verify the suitability of the 
truncation and to minimize errors induced by it.
Once an appropriate truncation scheme is established the Dyson-Schwinger framework 
offers a reliable  and robust tool to explore the theory at vastly different 
scales, and to acquire results
which might be too demanding to obtain from lattice simulations.

This paper is organized as follows. 
In section~\ref{sec:intro_coupled_system} we introduce the coupled
system of equations for the quark propagator and the Yang-Mills system, provide
details of the truncation as well as the employed renormalization scheme. We also
discuss general aspects concerning the IR behaviour of the system.
In section~\ref{sec:results} we present results obtained from the 
self-consistent treatment using chiral fermions on the one hand but also discuss 
the influence of non-vanishing fermion masses. 
Furthermore, we discuss the impact of the model parameters, where we also compare 
to lattice data in order to check the validity of the employed truncation scheme. 
We note, however, that our ansatz for the quark-gluon vertex is able to describe 
the system only on a qualitative level.
For a quantitative description the full quark-gluon vertex has to be included in 
the calculations which, however, goes beyond the scope of this paper. 
Nevertheless, investigations in this direction are ongoing, and an inclusion 
of vertex functions in self-consistent calculations 
in future work seems feasible. Finally, we conclude in
section~\ref{sec:conclusions}.
Technical aspects (concerning scale-setting, 
some variation of the employed vertex functions, and the treatment 
of spurious quadratic divergencies) can be found in three appendices.

%%%%%%%%%%%%%%%%%%%%%%%%%%%%%%%%%%%%%%%%%%%%%%%%%%%%%%%%%%%%%%%%%%%%%%%%%%%%%%%%%%%%%%%%%%%%%%%%%%%%%%%%%%%%%%%%%
\section{The System of Coupled Dyson-Schwinger Equations}
\label{sec:intro_coupled_system}

\begin{wrapfigure}{r}{0.5\columnwidth}
\center
\vspace{-0.8cm}
\includegraphics[width=0.5\columnwidth]{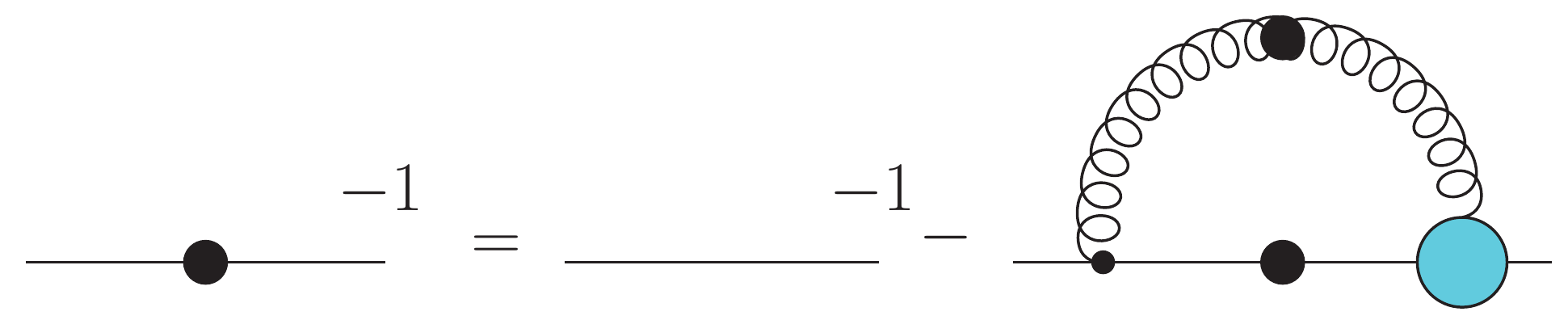}
\caption{The DSE for the quark propagator.}
\label{fig:quark_DSE}
\end{wrapfigure}

In the following investigation the central object is the DSE for the quark propagator 
depicted in figure~\ref{fig:quark_DSE}.
Here, the two crucial ingredients are the (dressed)
gluon propagator $D^{\mu\nu}(k)$ indicated by the wiggly line and 
the quark-gluon vertex $\varGamma^\nu(q,p;k)$ represented by the filled circle.
By increasing the number of flavours back-coupling effects of quark degrees of 
freedom on the Yang-Mills sector become more and more important. Thus, at 
some point simple model descriptions of the gluon propagator, and in particular, 
a naive extrapolation of QCD results \cite{Bashir:2013zha}, might not be sufficient
such that a self-consistent inclusion of the corresponding gluon DSE is a natural step.
In the following we introduce the coupled system of DSEs for the Yang-Mills 
and the matter sector as well as our truncation scheme for the quark-gluon vertex.
To keep the presentation reasonably self-contained  we give also an overview 
of the methods employed to solve the system self-consistently.

\subsection{The Quark Dyson-Schwinger Equation}
\label{sec:quark_DSE}
The renormalized DSE for the quark propagator is given by
 \begin{equation}
 \label{eq:quark_DSE}
 S^{-1}(p) = Z_2 S_0^{-1}(p) + g^2Z_{1F}C_F\int\frac{d^4q}{(2\pi)^4} 
 \gamma^\mu S(q)\varGamma^\nu(q,p;k)D^{\mu\nu}(k) \ ,
 \end{equation}
where we follow the conventions and notation of ref.~\cite{Fischer:2003rp}. 
$Z_2$ and $Z_{1F}$ are the quark wave function and the quark-gluon
vertex renormalization constants, respectively. 
From the colour trace one obtains a factor $C_F=(N_c^2-1)/(2N_c)$,
and the gluon momentum is defined as $k_\mu=p_\mu-q_\mu$.
The full quark propagator is given by
\begin{equation}
\label{eq:quark_propagator}
 S(p) = \dfrac{1}{-i\slashed p A(p^2,\mu^2)+B(p^2,\mu^2)} \ ,
\end{equation}
where the dressing functions $A$ and $B$ implicitly depend on the 
renormalization scale $\mu$.
The quark mass function $M(p^2)=B(p^2,\mu^2)/A(p^2,\mu^2)$ is a 
renormalization scale independent quantity.
The bare quark propagator is obtained via
$A(p^2,\mu^2)=1$ and $B(p^2,\mu^2)=m_0$, where, except for subsection~\ref{sec:quark_renormalization},
we omit in the following the explicit renormalization scale dependence of the dressing functions for brevity.

\begin{figure*}[!ht]
\center
\includegraphics[width=\columnwidth]{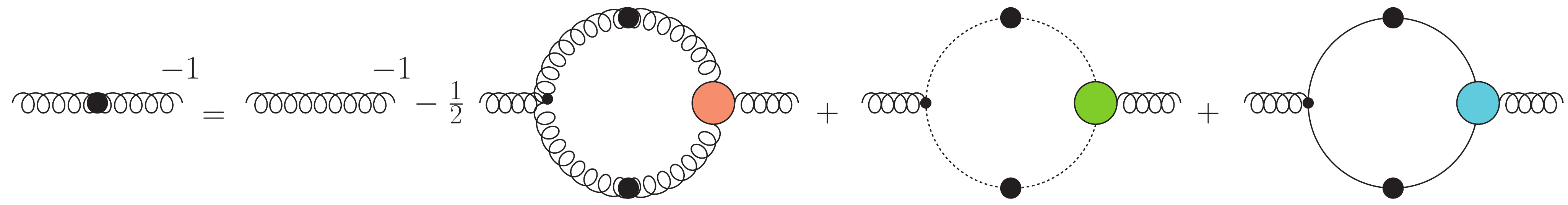} \\
\caption{The truncated DSE for the gluon propagator. Unquenching effects enter 
via the quark loop diagram in the last term of the equation, coupling the 
Yang-Mills part to the matter sector.}
\label{fig:gluon_DSE}
\end{figure*}

\begin{wrapfigure}{r}{0.5\columnwidth}
\center
\vspace{-0.8cm}
\includegraphics[width=0.5\columnwidth]{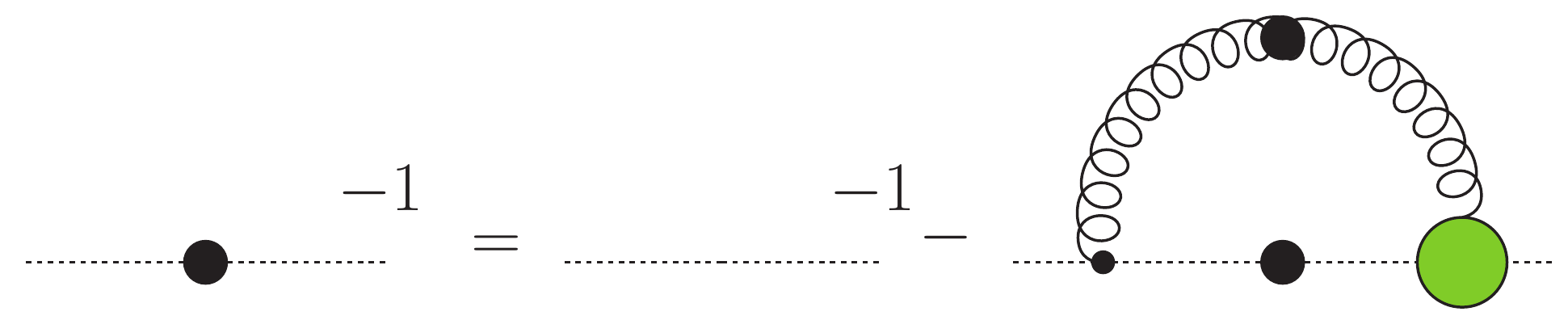}
\caption{The DSE for the ghost propagator.}
\label{fig:ghost_DSE}
\end{wrapfigure}

As detailed later in section~\ref{sec:gluon_equation} the gluon 
propagator $D^{\mu\nu}(p)$ 
is included self-consistently by solving the corresponding
DSEs for the Yang-Mills system depicted in 
figure~\ref{fig:gluon_DSE} and figure~\ref{fig:ghost_DSE}. 
Unquenching effects enter the gluon DSE via the quark-loop diagram.

A note on our truncation scheme is here  in order. (More details of the 
truncation can be found in refs.~\cite{Fischer:2003rp}.) 
Due to the considerable complications and technical obstacles induced by 
the treatment of four-gluon interactions in a self-consistent DSE approach 
as well as indications of their lack of importance \cite{Mader:2013ru} we 
neglect the two-loop diagrams  in the gluon equation. 
Furthermore, their contribution may be taken into account by adjusting the 
employed three-gluon vertex model as shown in ref.~\cite{Huber:2012kd}.
As we will detail below in section~\ref{sec:3gv_impact} the phase transition 
at $N_f^{crit}$ is quite insensitive to the 
details of the three-gluon vertex and the main impact seems to come from the 
different tensor structures of the full quark-gluon vertex.
Furthermore, for the ghost-gluon vertex a bare vertex approximation is sufficient, since
this object deviates only mildly from its tree-level structure, {\it cf. e.g.} 
refs.~\cite{Cucchieri:2004sq,Schleifenbaum:2004id,Cucchieri:2008qm,Huber:2012kd}.

Using this truncation scheme we obtain an almost closed system of equations
where the only missing ingredient is the quark-gluon vertex. 
Although this object was at the focus of earlier investigations 
\cite{Alkofer:2008tt,Hopfer:2013np,Williams:2014iea,Aguilar:2014lha}
it is up to now still too ambitious to include it in a full self-consistent way 
due to its quite complicated multi-tensor structure.
Thus, in order to proceed we defer this desirable but also highly demanding 
task to future work and adopt a model for the vertex which assumes
its factorisation into a non-Abelian scalar dressing function and 
an Abelian part that carries the tensor structure \cite{Fischer:2003rp}:
\begin{equation}
 \varGamma_\nu(p,q;k) = V_\nu^{Abel}(p,q;k)W^{\neg Abel}(p,q;k) \ .
\label{eq:qgv_model}
\end{equation}

Some motivation for this ansatz can be gained from the Slavnov-Taylor
identity of the vertex \cite{Marciano:1977su}. It is given by  
\begin{eqnarray}
i\: k_\mu \: \Gamma_\mu(q,k) = G(k^2) [S^{-1}(p) H(p,q) - \bar{H}(q,p) \: S^{-1}(q)] \ ,
\label{quark-gluon-STI}
\end{eqnarray}
with ghost dressing function $G(k^2)$ and ghost-quark scattering kernel $H(q,p)$ 
together with 'conjugate' $\bar{H}$. Furthermore, $p,q$ are the momenta of the two 
quarks and $k=p-q$ is the corresponding gluon momentum. Since the
non-perturbative behavior of $H(p,q)$ and its conjugate is currently unknown,
there is no exact solution of this identity available. However, some general 
structural information can be gained. Consider for the moment $H=\bar{H}=1$.
Then, using the additional requirement of regularity at zero gluon momentum,
this approximate STI can be solved exactly by a multiplicative ansatz given by
the Ball-Chiu vertex (which solves the Abelian version of this identity \cite{Ball:1980ay}) 
and an extra multiplicative factor $G(k^2)$, which provides for a non-Abelian enhancement 
of the vertex at small momenta. This observation is the main motivation behind an
ansatz such as the one given in eq.~(\ref{eq:qgv_model}). 

What is then the effect of the scattering kernel $H(q,p)$ in the STI? First, it 
is known from approximate solutions for $H$, that the scattering kernel provides for 
additional infrared enhancement \cite{Aguilar:2013ac,Rojas:2013tza}. Second, it is well-known
that the combined effect of the gluon dressing and that of the quark-gluon vertex in the 
ultraviolet momentum region has to resemble the running coupling of QCD, otherwise
resummed perturbation theory cannot be reproduced by the quark-DSE. An ansatz for 
the vertex that provides both, further infrared enhancement and resummed perturbation theory 
is given by \cite{Fischer:2003rp}
\begin{equation}
W^{\neg Abel}(p,q;k) = G^2(k^2) \:\tilde{Z}_3 \ ,
\label{eq:vertex_nonabel}
\end{equation}
where $\tilde{Z}_3$ is the ghost renormalization factor. We use this ansatz in the following.

For the Abelian part of the
vertex we employ the leading term in the Ball-Chiu construction, i.e.
\begin{equation}
V_\nu^{abel}(p,q;k) = \frac{A(p^2)+A(q^2)}{2} \gamma_\nu \ ,
\label{eq:vertex_abel_1BC}
\end{equation}
for the calculations presented in the main part of this work. We have also used 
the full Curtis-Pennington construction including all terms of the Ball-Chiu vertex 
plus an additional transverse part. Since the corresponding calculation is much more 
involved (with larger numerical errors) but did not change our results qualitatively,
we present only some results in appendix \ref{sec:vertex_impact}. Clearly, the stability 
of the qualitative features of our results with respect to changes in the Abelian part 
of the vertex serves as an indication for the robustness of our results beyond the simple 
vertex model we use. This is further discussed in the appendix.

\subsection{The Renormalization Scheme}
\label{sec:quark_renormalization}
In the following we use a MOM renormalization scheme \cite{Fischer:2003rp}.
The dressing functions for the quark propagator can formally be written as
\begin{eqnarray}
                      A(p^2,\mu^2) & = & Z_2 + Z_{1F}\,\Pi_A(p^2,\mu^2) \ , \label{eq:MOM_A_equation} \\
 B(p^2,\mu^2) = M(p^2)A(p^2,\mu^2) & = & Z_2 m_0 + Z_{1F}\,\Pi_M(p^2,\mu^2) \ ,\label{eq:MOM_B_equation} 
\end{eqnarray}
where we omit the explicit renormalization scale and cutoff dependence of the renormalization 
constants for brevity, {\it i.e.} it is understood that $Z_i=Z_i(\mu^2,\Lambda^2)$.
Furthermore, $Z_2$ and $Z_{1F}$ are connected to the ghost renormalization 
constant $\tilde Z_3$ by a Slavnov-Taylor identity via the relation
$Z_{1F} = \tilde Z_1 Z_2/\tilde Z_3 = Z_2/\tilde Z_3$ \cite{Marciano:1977su},
where we used $\tilde Z_1=1$ for the ghost-gluon vertex renormalization constant 
since this object is UV finite in Landau gauge~\cite{Taylor:1971ff}.\footnote{The momentum subtraction
of the propagators together with $\tilde Z_1=1$ defines the so-called MiniMOM 
scheme~\cite{vonSmekal:2009ae}, {\it i.e.} no further renormalization conditions
for the vertex functions are required, {\it cf.} also \cite{Lerche:2002ep}.}
Furthermore, the bare quark mass $m_0(\Lambda^2)$ is related to the renormalized mass 
$m_R(\mu^2)$ by $m_0(\Lambda^2)=Z_m(\mu^2,\Lambda^2) m_R(\mu^2)$, 
where $Z_m$ denotes the mass renormalization constant and $\Lambda^2$ is the squared cutoff.
By rewriting eq.~\eqref{eq:MOM_A_equation} and using the explicit occurrence of $\tilde Z_3$ in 
eq.~\eqref{eq:vertex_nonabel} we get
\begin{equation}
\label{eq:MOM_scheme_Z2}
 Z_2^{-1} = A^{-1}(p^2,\mu^2)\Bigl[1+\Pi_A(p^2,\mu^2)\Bigr] \ ,
\end{equation}
where subsequently we apply a MOM scheme such that the vector self-energy $A(p^2,\mu^2)$ is written as
\begin{equation}
\label{eq:MOM_scheme_A}
 A(p^2,\mu^2) = \frac{1+\Pi_A(p^2,\mu^2)}{1+\Pi_A(\mu^2,\mu^2)} \ ,
\end{equation} 
with the renormalization condition $A(\mu^2,\mu^2)=1$.
From eq.~\eqref{eq:MOM_B_equation} we analogously obtain
\begin{equation}
\label{eq:MOM_scheme_B}
 B(p^2,\mu^2) = m_R(\mu^2) + Z_2\,\Bigl[\Pi_M(p^2,\mu^2) - \Pi_M(\mu^2,\mu^2)\Bigr] \ ,
\end{equation} 
where we made use of the identity $M(\mu^2)=m_R(\mu^2)$ for the quark mass function $M$ and the renormalized mass $m_R$ 
which is valid for perturbative renormalization scales $\mu^2$. 
In the chiral limit eq.~\eqref{eq:MOM_scheme_B} simplifies to $B(p^2,\mu^2) = Z_2\,\Pi_M(p^2,\mu^2)$,
where $Z_2$ is obtained from eq.~\eqref{eq:MOM_scheme_Z2} within each iteration step.
The evolution of the perturbative mass $m_R(\mu^2)$ to another perturbative scale $\nu^2$ 
is performed via
\begin{equation}
 m_R(\nu^2) = m_R(\mu^2) \left(\dfrac{\ln(\mu^2/\Lambda_{QCD}^2)}{\ln(\nu^2/\Lambda_{QCD}^2)}\right)^{\gamma_m} \ ,
\end{equation}
where $\gamma_m=12/(11N_c-2N_f)$ denotes the anomalous dimension of the quark mass function.

\subsection{Order Parameters and Scale Setting}
\label{sec:orderparameters}
From the quark propagator \eqref{eq:quark_propagator} different quantities  can be
extracted which are suitable to study the phase transition. Here, the quark mass
function $M(p^2)$ in the limit of  vanishing momenta  is the most
straightforward order parameter to investigate. Closely related is the chiral
condensate obtained from integrating the quark propagator in the chiral limit
\begin{equation}
 \langle\bar\psi\psi\rangle_\mu=Z_2 Z_m N_c \int\frac{d^4q}{(2\pi)^4}\, {\mathrm tr}_D S(q) \ ,
\end{equation}
where its renormalization scale independent form can be calculated via
\begin{equation}
 \label{eq:chiral_condensate}
 \langle\bar\psi\psi\rangle = \left(\frac{1}{2}\ln(\mu^2/\Lambda_{QCD}^2)\right)^{-\gamma_m}\langle\bar\psi\psi\rangle_\mu \ .
\end{equation}
Furthermore, for the pion decay constant we use the approximation \cite{Pagels:1979hd}
\begin{equation}
 \label{eq:PDC}
 f_\pi^2 = \dfrac{N_c}{4\pi^2}Z_2
 \int dq^2\,q^2\dfrac{M(q^2)A^{-1}(q^2)}{(q^2+M^2(q^2))^2}
 \left(M(q^2)+\dfrac{q^2}{2}M'(q^2)\right) \ .
\end{equation}
As detailed in appendix~\ref{sec:scale_fixing} this quantity is employed to fix 
the physical scale of the system.

\subsection{The Gluon Dyson-Schwinger Equation}
\label{sec:gluon_equation}

\begin{wrapfigure}{r}{0.4\columnwidth}
\center
\vspace{-0.8cm}
\includegraphics[width=0.4\columnwidth]{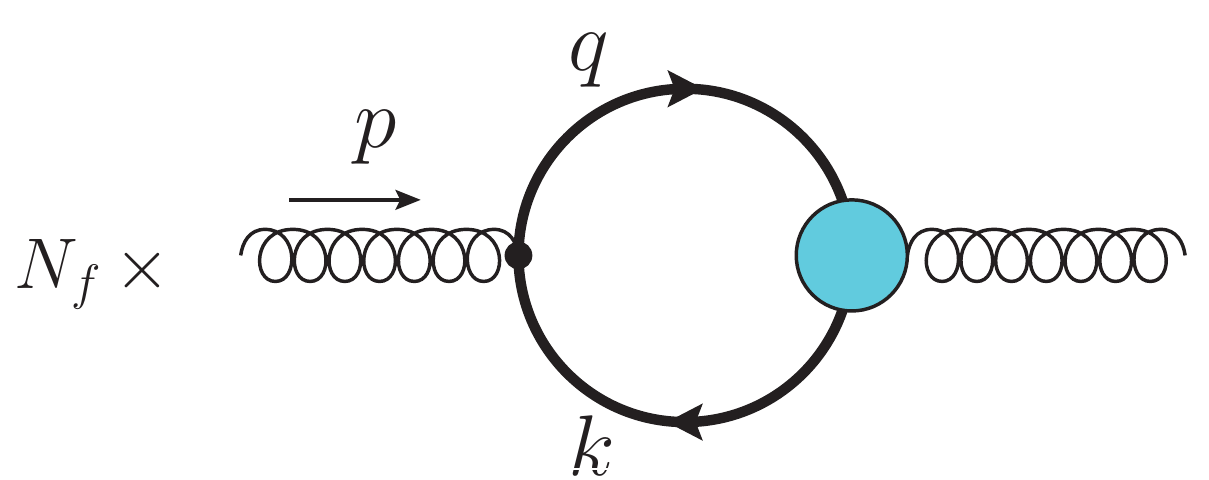}
\caption{The quark loop diagram entering the DSE for the gluon propagator. $N_f$ is the number of flavours.}
\label{fig:quark_loop}
\end{wrapfigure}

In order to study the transition to the chirally symmetric phase at a large number of flavours, the DSE for the quark propagator
together with the Yang-Mills system has to be solved self-consistently. For details on the numerical implementation we
refer the reader to refs.~\cite{Fischer:2003rp,Fischer:2005en,Hopfer:2012ht}.
The formal structure of the gluon DSE is given by
\begin{equation*}
 D_{\mu\nu}^{-1}(p) = Z_3 D_{0,\mu\nu}^{-1}(p) + \Pi_{\mu\nu}^{YM}(p) + \Pi_{\mu\nu}^{quark}(p) \ .
\end{equation*}
Unquenching effects enter this equation via the last term.
To be more precise, the gluon self-energy contribution stemming from the quark-loop is given by
\begin{equation}
 \Pi_{\mu\nu}^{quark}(p) = - g^2 \frac{N_f}{2}Z_{1F}\int\frac{d^4q}{(2\pi)^4}\, 
 tr_D\Bigl[\gamma_\mu S(q)\varGamma_\nu(q,k;p)S(k)\Bigr] \ ,
 \label{eq:quarkloop_SE}
\end{equation}
where details on the momentum routing are shown in figure~\ref{fig:quark_loop}. 
For the quark-gluon vertex in the quark loop we use a symmetric version of the non-Abelian model 
presented in eq.~\eqref{eq:vertex_nonabel} which takes the form
\begin{equation}
W^{\neg Abel}(q,k;p) = G(q^2) \,G(k^2) \:\tilde{Z}_3 \ ,
\label{eq:vertex_quarkloop_nonabel}
\end{equation}
with quark momenta $q$ and $k$. This change of arguments as compared to eq.~(\ref{eq:vertex_nonabel}) 
can be justified as follows \cite{Fischer:2003rp}: First, the kinematical structure of the 
quark-loop in the gluon-DSE is such, that different kinematical sections of the full quark-gluon
vertex are probed in the calculation. Thus an ansatz that is good in the quark-DSE may be significantly
worse in the gauge sector of the theory. This is discussed e.g. in ref.~\cite{Kizilersu:2009kg}.
Second, one observes, that multiplicative dressing functions in the vertex that depend on the gluon 
momentum only destroy the QCD property of multiplicative renormalisability of the gluon-DSE. Taken
together, these two observations motivate the change of arguments in eq.~(\ref{eq:vertex_quarkloop_nonabel}).

Contracting eq.~\eqref{eq:quarkloop_SE} with the transverse projector 
$ \mathcal{P}_{\mu\nu}(p) = \delta_{\mu\nu} - p_\mu p_\nu/p^2$
one obtains 
\begin{eqnarray}
\Pi_{quark}(p^2) &=& -g^2 N_f Z_2 \int \frac{d^4q}{(2\pi)^4}\, \sigma_v(q^2)\sigma_v(k^2)G(q^2)G(k^2) \nonumber \\
&&\qquad\qquad\qquad\qquad\times\frac{A(q^2)+A(k^2)}{2}\,U(p^2,q^2,k^2) \ ,
\label{eq:quarkloop_SE_contracted}
\end{eqnarray}
where we abbreviated the vector part of the quark propagator by $\sigma_v(p^2) \equiv A^{-1}(p^2)/(p^2+M^2(p^2))$.
The kernel $U$ is given by (note that a color factor $1/2$ as well as 
a factor of $1/3p^2$ from the left hand side of the gluon equation has been absorbed):
\begin{eqnarray}
U(p^2,q^2,k^2) &=&\frac{k^4}{3 p^4} + \frac{k^2}{p^2}\left(\frac{1}{3} - \frac{2 q^2}{3p^2}\right)
             -\frac{2}{3}+\frac{q^2}{3p^2}  + \frac{q^4}{3p^4} \ . \label{eq:U}
\end{eqnarray}

\begin{figure}[t]
\center
\includegraphics[width=0.6\columnwidth]{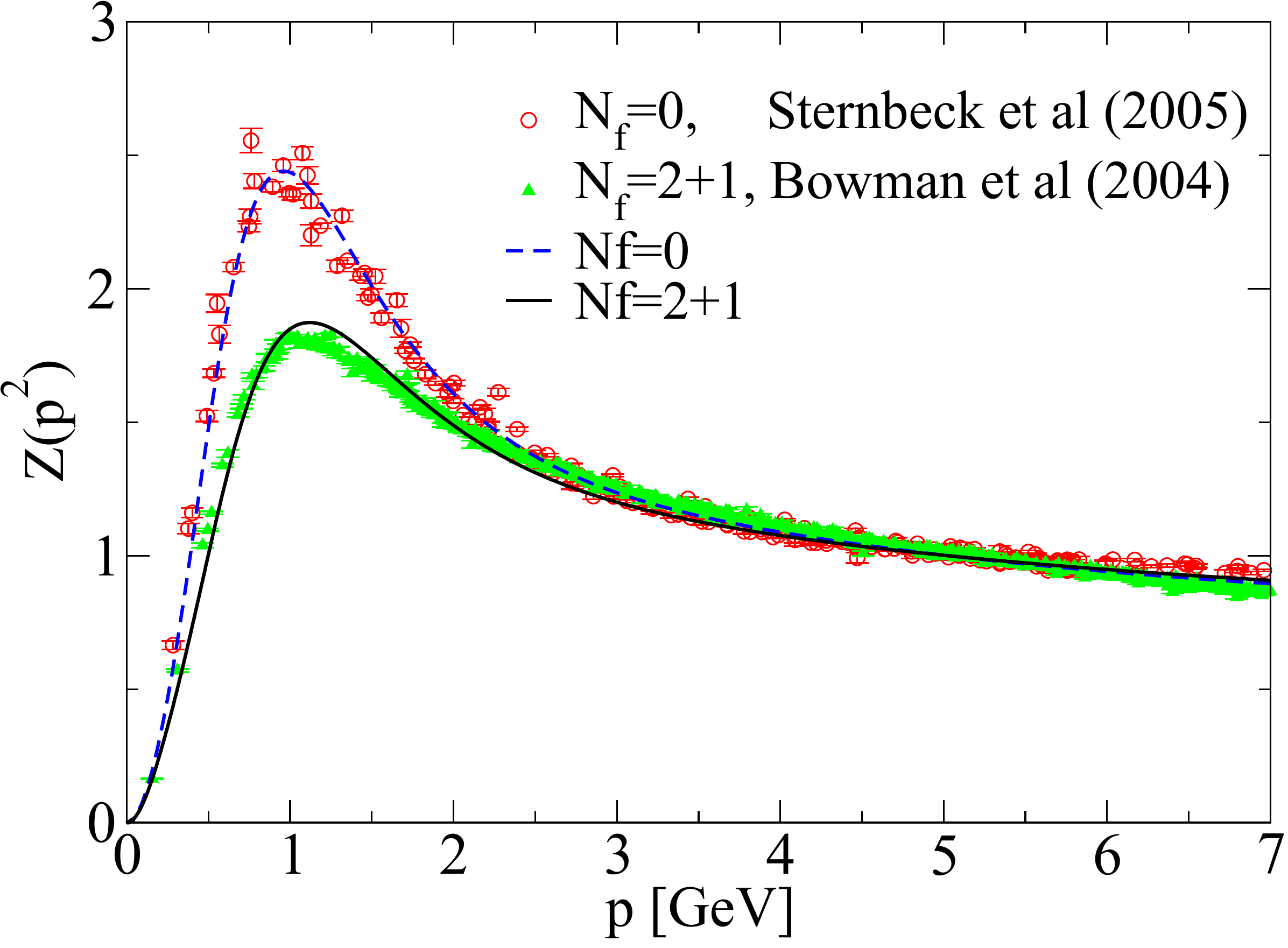}
\caption{A comparison of our DSE result for the gluon dressing function with lattice data
for both the quenched and the unquenched case. We find very good agreement in both cases. }
\label{fig:lattice_comparison}
\end{figure}

Using the ansatz for the three-gluon vertex proposed in ref.~\cite{Huber:2012kd} and 
our truncation for the quark-gluon vertex from above, we first solved the coupled
system of equations for $N_f=0$ and $N_f=2+1$ flavours.
In figure~\ref{fig:lattice_comparison} we show results for the gluon dressing function
obtained from our calculations and compare with corresponding lattice data.
We find very good agreement for both, the quenched and the unquenched case.
Thus our ansatz is well-suited to describe the system at least on a qualitative
level. However, we also note that eq.~\eqref{eq:qgv_model} is certainly not 
sufficient to capture the full physics inherent in the vertex due to the missing 
tensor structures. This issue will be addressed in more detail below and in 
appendix \ref{sec:qgv_impact}.

\section{Numerical Results for Propagators and Running Coupling}
\label{sec:results}
We present first results obtained with the ansatz for the quark-gluon vertex
defined in eq.~\eqref{eq:vertex_nonabel} and eq.~\eqref{eq:vertex_quarkloop_nonabel}
(dubbed  $1BC\times G^2$ ansatz in the following). For the gauge-boson vertex the model
\eqref{eq:3gv_Fischer} is employed.
\begin{figure}[!b]
\center
\subfigure{\includegraphics[width=0.49\columnwidth]{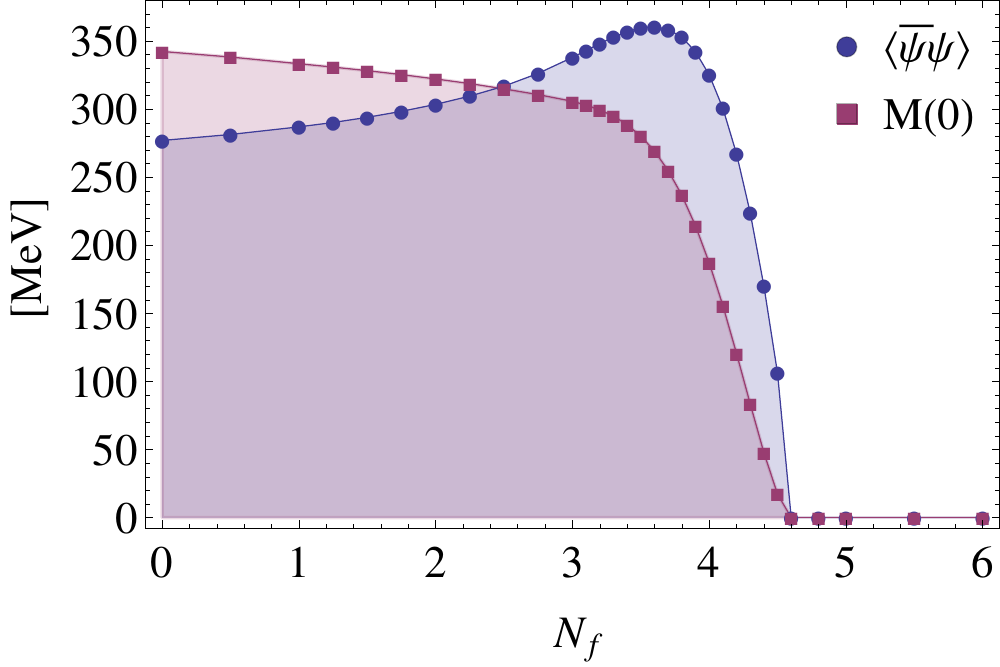}\label{fig:1BCxG2_Results_Mass}\hspace*{0.1cm}}
\subfigure{\includegraphics[width=0.49\columnwidth]{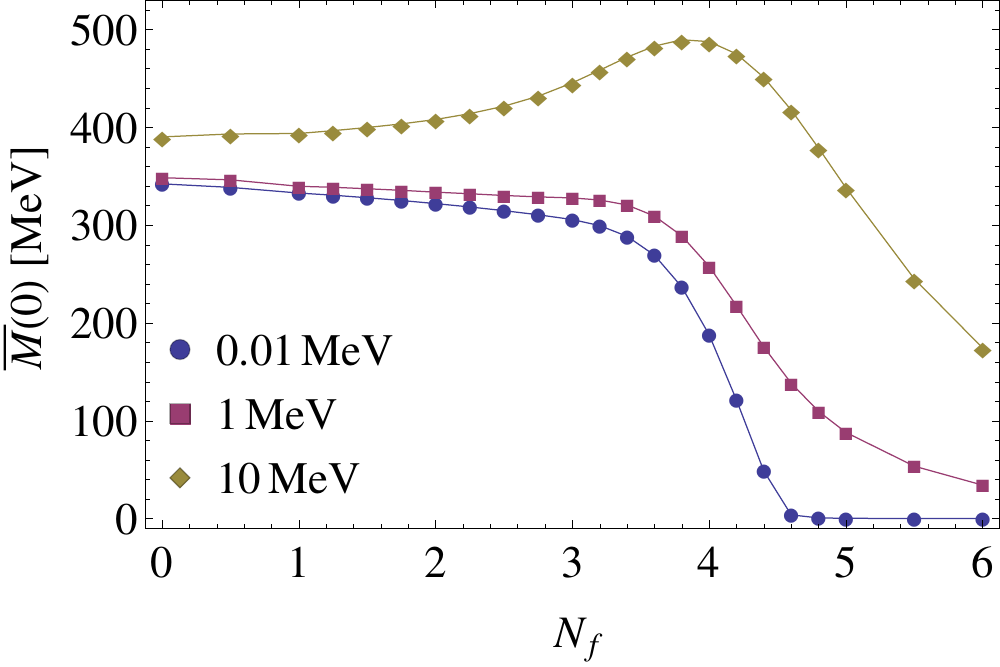}\label{fig:1BCxG2_Results_Mass_Dependence}}
\caption{The quark mass function $M(p^2)$ in the limit of vanishing external momentum $p^2$
as well as the chiral condensate $\langle\bar\psi\psi\rangle$ for 
different flavour numbers $N_f$. Lines are drawn to guide the eye.
Left (a): Calculations performed in the chiral limit. 
For $N_f>N_f^{crit} \approx 4.5$ no dynamical mass is  generated.
Right (b): Results obtained with different bare quark masses. 
$\overline M(0)$ is defined as the dynamically generated
infrared quark mass minus the bare quark mass 
specified at 2 GeV, {\it i.e}, $\overline M(0)=M(0)-m_0@\,2$ GeV. 
As expected the phase transition changes to a crossover.}
\end{figure}
Using this truncation the critical number of flavours is 
$N_f^{crit} \approx 4.5$ as shown
in figure~\ref{fig:1BCxG2_Results_Mass}, 
where we plot the quark mass function $M(p^2\to 0)$
as well as the chiral condensate~\eqref{eq:chiral_condensate} 
versus the number of flavours $N_f$. For $N_f=5$ no dynamical mass is generated, and one obtains 
a chirally symmetric phase. 

Compared to previous results from the functional
renormalization group \cite{Braun:2009ns,Braun:2010qs} our number is much smaller
and indeed seems unnaturally small. We attribute this to the differences in the 
truncation schemes, which need to be studied in more detail in future work. 
However, as already discussed above, our study is not so much concerned 
with the actual number for $N_f^{crit}$ but with the qualitative behaviour of 
the propagators and in particular with the running coupling in the symmetric
phase above $N_f^{crit}$. We are confident that our approximation is robust enough 
to allow for meaningful results in this respect. 

Furthermore, we investigate the impact of finite bare quark masses on the system. 
In figure~\ref{fig:1BCxG2_Results_Mass_Dependence}
we show results for the quark mass function using different bare quark masses 
as input.
As expected the explicit conformal symmetry breaking leads to a crossover
at the same value of $N_f^{crit}$ where the quark mass function in the 
chiral limit dropped to zero.

\begin{figure*}[!b]
\center
\subfigure{\includegraphics[width=0.49\columnwidth]{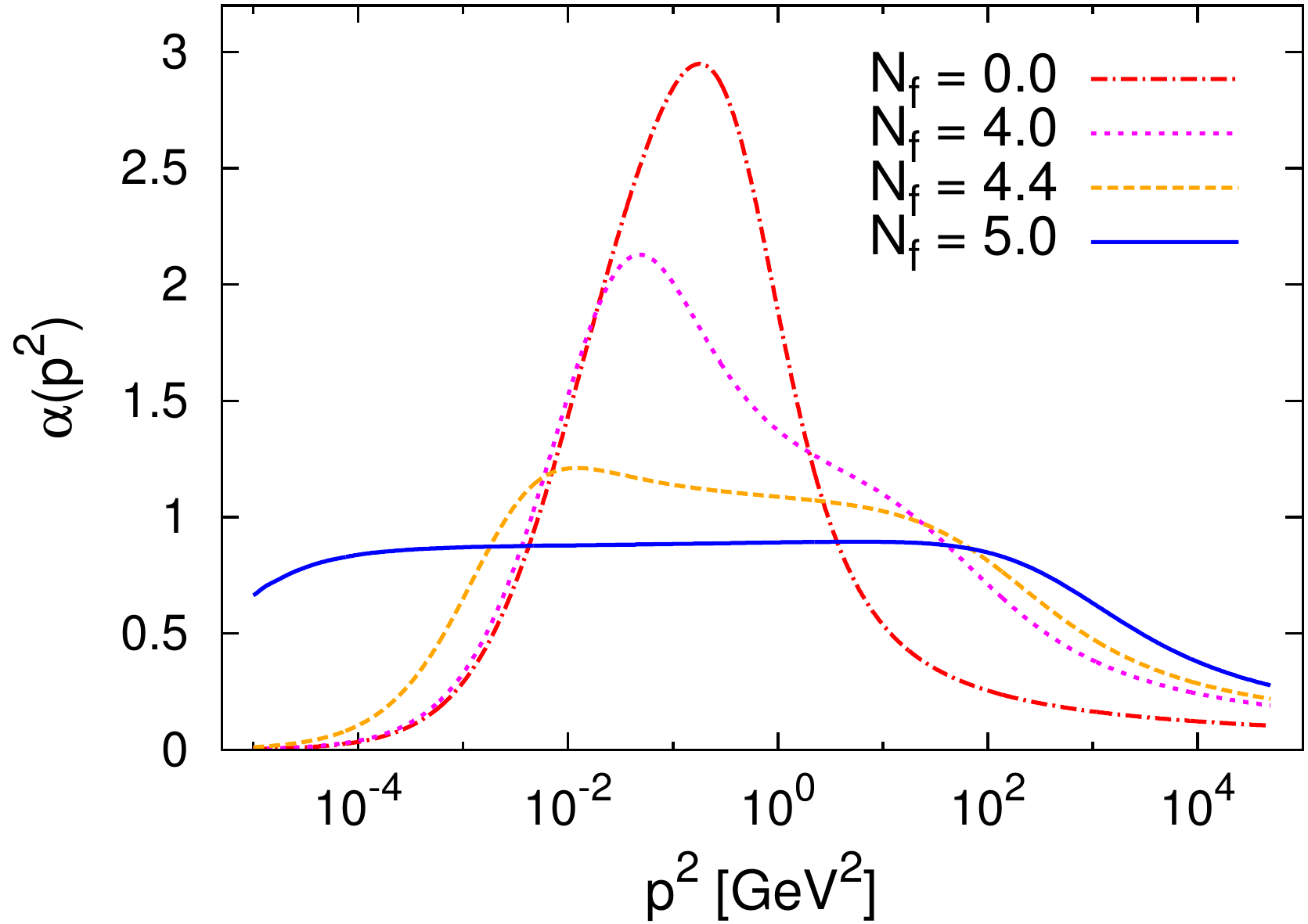}\label{fig:1BCxG2_Results_coupling}}
\subfigure{\includegraphics[width=0.49\columnwidth]{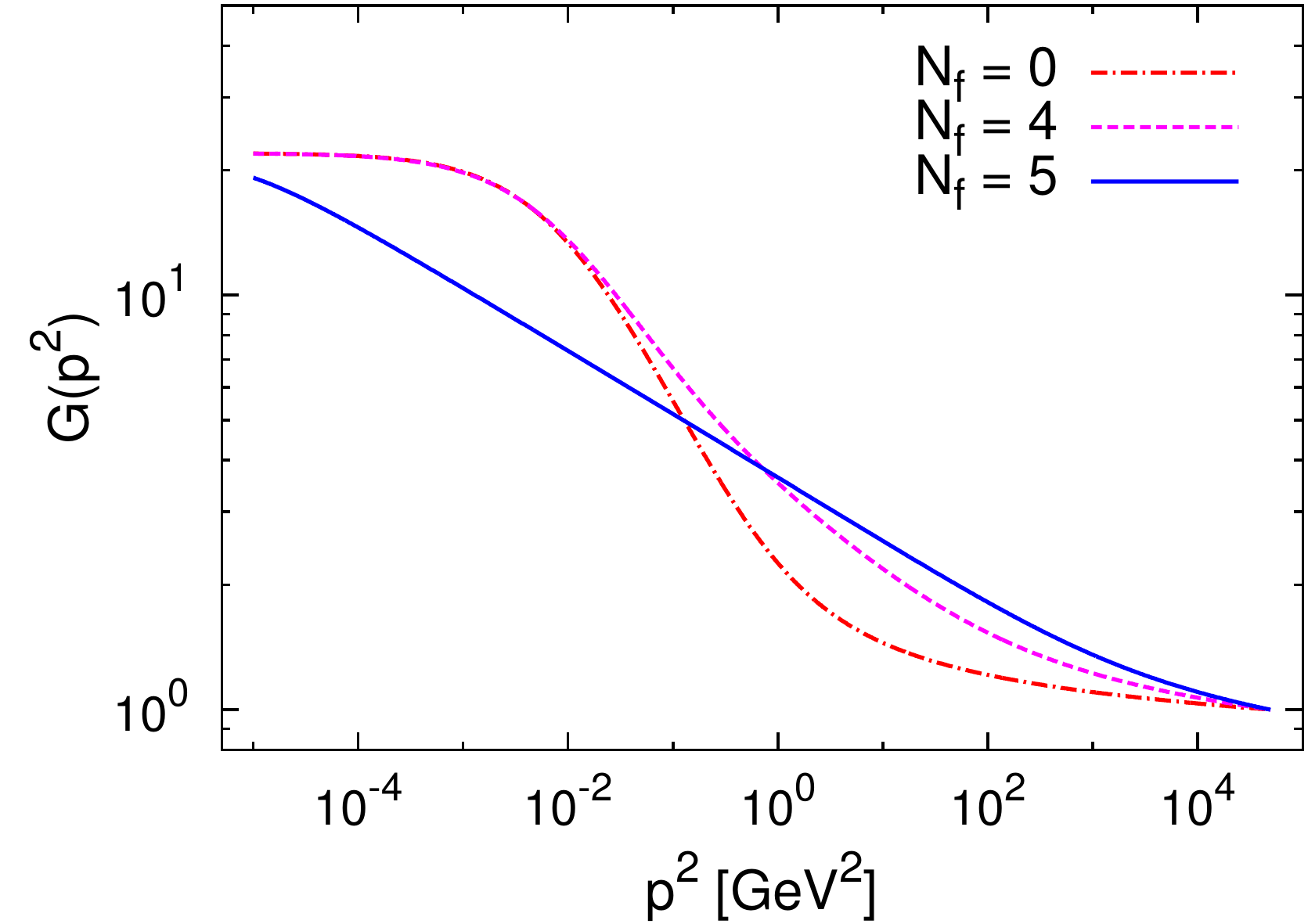}\label{fig:1BCxG2_Results_ghost}} \\
\subfigure{\includegraphics[width=0.49\columnwidth]{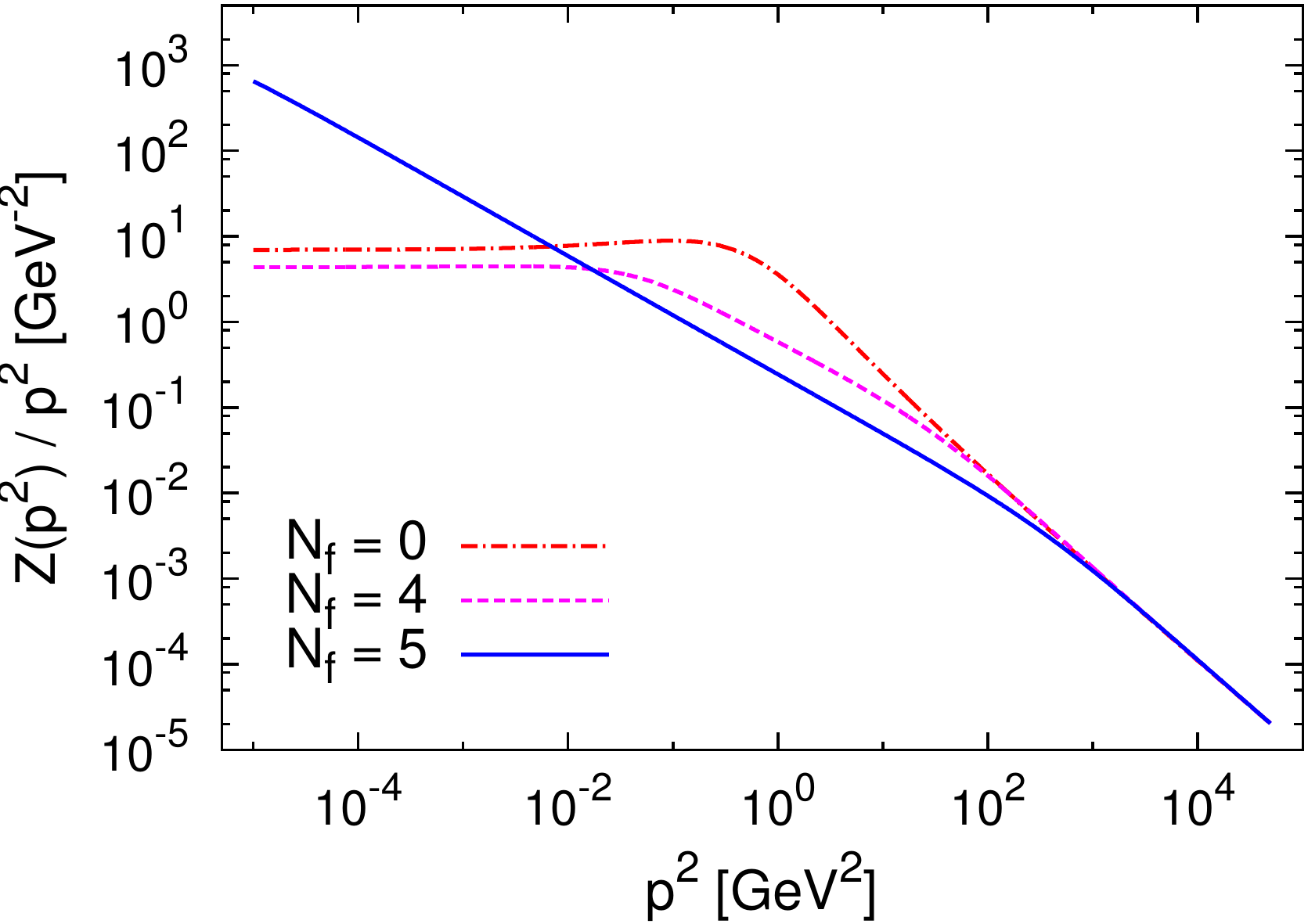}\label{fig:1BCxG2_Results_gluon}}
\subfigure{\includegraphics[width=0.49\columnwidth]{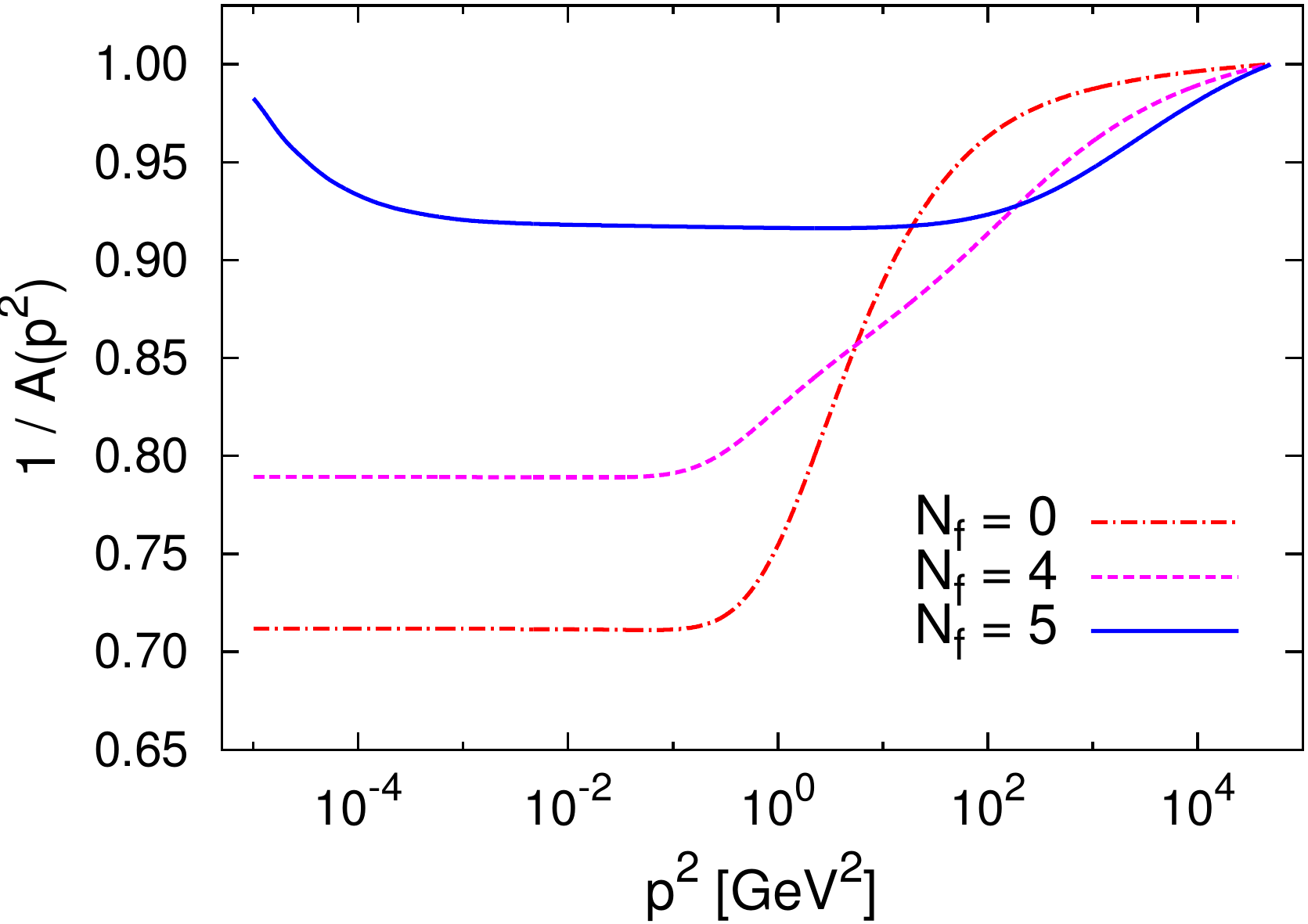}\label{fig:1BCxG2_Results_A}}
\caption{Results for (a) the running coupling $\alpha(p^2)$ (upper left panel), 
(b) the ghost dressing function $G(p^2)$ (upper right panel), 
(c) the gluon propagator $Z(p^2)/p^2$ (lower left panel), and 
(d) the quark wave-function renormalization $A^{-1}(p^2)$ (lower right panel) 
using a $1BC\times G^2$ ansatz for the quark-gluon vertex. 
We use a perturbative renormalization scale of $\mu^2=5\times 10^4$ GeV$^2$.}
\label{fig:1BCxG2_Results}
\end{figure*}

In figure~\ref{fig:1BCxG2_Results} we present results for the running coupling 
$\alpha(p^2)=\alpha(\mu^2)Z(p^2)G^2(p^2)$,
the ghost dressing function $G(p^2)$, the gluon propagator $Z(p^2)/p^2$ and the 
inverse vector self-energy $A^{-1}(p^2)$ for different flavour values $N_f\in\{0,4,5\}$.
By increasing the number of flavours the coupling is lowered where already at
$N_f\lesssim N_f^{crit}$ this decrease is significant. For $N_f=4$ the onset of 
the building of a plateau is visible, which becomes flatter with further increase
of $N_f$ until it develops a large momentum range with zero gradient at $N_f^{crit}$.
If $N_f$ is further increased the height of this plateau is successively
lowered. This behaviour can be traced back to a scaling relation between the
ghost and gluon  propagators in a large momentum region within the 
chirally symmetric phase. One clearly identifies a power law behaviour
as can be seen from figure~\ref{fig:1BCxG2_Results_ghost} and
 figure~\ref{fig:1BCxG2_Results_gluon}.
Indeed, in this chirally symmetric phase the system can be treated analytically
using power law ansaetze for the propagators since the mass term
in the vector part of the quark propagator $\sigma_v$ vanishes, 
{\it cf. e.g.} ref.~\cite{Lerche:2002ep} for a corresponding treatment of the pure Yang-Mills system.
Using the ansaetze \eqref{eq:vertex_nonabel} and \eqref{eq:vertex_quarkloop_nonabel}
for the quark-gluon vertex and the ghost boundary condition $G^{-1}(0)\rightarrow 0$
one is able to find a scaling relation for the Yang-Mills
dressing functions given by 
\begin{equation}
Z(p^2)\propto (p^2)^{2\varrho} \ , \hspace*{5mm} G(p^2)\propto (p^2)^{-\varrho} \ ,
\end{equation}
which together result in the plateau behaviour of the running coupling seen in
fig.~\ref{fig:1BCxG2_Results_coupling}. 
The quark dressing function $A(p^2)\propto const$ in this case which also 
agrees with our numerical results. For the $N_f$-dependent exponent $\varrho$ one obtains 
$\varrho\approx 0.15$ at $N_f^{crit}$. Whereas the qualitative behaviour of the
ghost and gluon dressing functions together with the 
flatness of the running coupling
may very well be robust with respect to changes in the truncation scheme (see below), 
we regard the value for $\varrho$ as tentative only.
The small value of the running coupling relates to the behaviour of the quark 
mass function in figure~\ref{fig:1BCxG2_Results_Mass}, the resulting interaction
is not strong enough to generate mass dynamically.

Within this study the question of the impact of the employed vertex models onto
the results is of uttermost importance. Therefore we studied the system of
propagator DSEs with different models for  the three-gluon and quark-gluon vertex
functions. As  the results are very similar to the ones displayed in 
fig.\ \ref{fig:1BCxG2_Results} (with one interesting exception discussed below)
we present them only in appendix~\ref{sec:vertex_impact}. The conclusion of this  comparison is that
neither the tested changes in the three-gluon vertex nor some changes in the
leading tensor structure of the quark-gluon vertex modifies our results 
significantly,  especially one sees the same type of behaviour for the
propagators  and the value for $N_f^{crit}$ does not 
change\footnote{A further technical subtlety is the correct treatment of spurious
quadratic divergencies in the gluon DSE. Therefore we also compare numerical
results obtained with different employed subtraction methods in
appendix~\ref{app:qDvgs_methods}, see figure~\ref{fig:qDvgs_subtraction_results}.
In general all methods yield the same results, where for small flavour numbers  we
even find excellent agreement within numerical accuracy. However, if the system comes
close to the phase transition some small deviations are observed. Nevertheless,
one can conclude that the spurious divergencies are safely removed.}. 

In order to test the influence of the multi-tensor structure of the quark-gluon 
vertex we generalize its Abelian part \eqref{eq:vertex_abel_1BC} to an ansatz 
originally used in QED \cite{Curtis:1990zs}, the so-called Curtis-Pennington vertex.
Besides containing all tensors of the Ball-Chiu vertex it includes one purely transverse 
term with a chirally even and a chirally odd part. The original motivation for 
constructing this vertex contribution was to restore multiplicative renormalizibility. 
Here we simply use it as a test for estimating the influence of beyond-tree-level 
tensor structures on our results within a numerically not too demanding setting. 
Whereas the behaviour of the propagators and the running coupling comes out extremely 
similar to the cases discussed above, and this below as well as above the phase 
transition, the value of $N_f^{crit}$ increases significantly to  $N_f^{crit} \approx 5.3$.
It is beyond the scope of this paper to include all possible eight transverse tensor 
structures, especially as a significant improvement of the employed truncation would 
require a self-consistent update of the quark-gluon vertex together with the solution
of the propagators. However, it is plain from the observed shift of $N_f^{crit}$ that such 
a more complete calculation may lead to a larger value for $N_f^{crit}$ closer
to the results from the functional renormalization group discussed above 
\cite{Braun:2009ns,Braun:2010qs}. On 
the other hand, the robustness of the results for the propagators provide substantial 
evidence that a very similar behaviour for them will be found in a more 
complete calculation. 

\section{Conclusions}
\label{sec:conclusions}

Based on models for the three-gluon and quark-gluon vertex functions a self-consistent
calculation of the Landau gauge 
propagators and the running coupling in a QCD-like gauge theory with an increased number
of light, resp., massless quark flavours has been presented. The main result is: 
The behaviour of these propagators and the running coupling change drastically when
entering the conformal window. This is obvious for the quark propagator because chiral 
symmetry is then not dynamically broken any more, and correspondingly no mass is 
dynamically generated. However, and this is the more important result, it is also true for
the gluon and ghost propagators. Whereas, in the chiral limit, the quark propagator hardly
deviates from the one of a free massless Dirac fermion, the gluon and ghost propagators
assume non-trivial power laws interrelated by a scaling relation 
such that the running coupling calculated from these
propagators stays constant over more than six orders of magnitude in $p^2$. 

This calculation also sheds some light on the reason
why a conformal window exists: For a small number
of light flavours the antiscreening of the gluons wins on all scales 
over the screening caused by quarks. For momenta smaller than the ones in the 
perturbative regime chiral symmetry is dynamically broken, a quark mass is generated, and
the quarks decouple. In the 
opposite end of a very large number of light flavours the screening of quarks wins against
the anti-screening of gluons on all scales. Asymptotic freedom is lost, and the theory is 
likely to be trivial from a Renormalization Group perspective. Now, for an intermediate
number of light flavours in the far UV the anti-screening of gluons is still dominant.
When lowering the scale the running coupling increases but now, with the enhanced
screening caused by quark loops,  the coupling never exceeds the critical value needed for
dynamical chiral symmetry breaking. Therefore no quark mass is generated and the quarks do
not decouple: They are long-range and dominate over the gluons at non-perturbative scales. 
Therefore, the quark loop is the driving term in the gluon equation, and 
its net effect is a freezing of the coupling.

Naturally, the question about the fate of confinement in the
conformal window arises. To this end, it is interesting to note that the gluon propagator
features no maximum any more. This is a clear indication that the violation of positivity
for transverse gluons (undoubtedly present in the confining phase) may cease to exist when
increasing the number of flavours above the critical one.

Although we are convinced that the qualitative conclusions presented here are robust
against an improvement of the employed vertex functions it is plain from the performed
comparisons of different quark-gluon vertex models that quantitative predictions are only
possible if (at least) the quark-gluon vertex is determined also self-consistently 
together with the propagators. Such an investigation is technically demanding but
nevertheless subject of on-going investigations. 

%%%%%%%%%%%%%%%%%%%%%%%%%%%%%%%%%%%%%%%%%%%%%%%%%%%%%%%%%%%%%%%%%%%%%%%%%%%%%%%%%%%%%%%%%%%%%%%%%%%%%%%%%%%%%%%%%
\begin{acknowledgments}
We thank Markus Huber, Axel Maas, Valentin Mader, Francesco Sannino and Milan Vujinovic for discussions.
MH acknowledges support from the Doktoratskolleg ''Hadrons in Vacuum, Nuclei and Stars`` of the 
Austrian Science Fund, FWF DK W1203-N16.
\end{acknowledgments}  
%%%%%%%%%%%%%%%%%%%%%%%%%%%%%%%%%%%%%%%%%%%%%%%%%%%%%%%%%%%%%%%%%%%%%%%%%%%%%%%%%%%%%%%%%%%%%%%%%%%%%%%%%%%%%%%%%

\appendix
\section{Scale Setting}
\label{sec:scale_fixing}
After applying a MOM renormalization scheme the DSEs for the Yang-Mills system read
\begin{eqnarray}
 G(p^2)^{-1} & = & \tilde Z_3 + \varPi_G(p^2) \quad\stackrel{MOM}{\rightarrow}\quad G(p^2)^{-1} = G(p^2_G)^{-1} + \varPi_G(p^2) - \varPi_G(p^2_G) \ , \qquad \\
 Z(p^2)^{-1} & = & Z_3 + \varPi_Z(p^2) \quad\stackrel{MOM}{\rightarrow}\quad Z(p^2)^{-1} = Z(p^2_Z)^{-1} + \varPi_Z(p^2) - \varPi_Z(p^2_Z) \ .
\end{eqnarray}
Thus, instead of dealing with the renormalization constants $Z_3$ and $\tilde Z_3$ one has to specify
the boundary conditions $G(p^2_G)^{-1}$ and $Z(p^2_Z)^{-1}$. It is numerically 
convenient to use the subtraction points $p^2_G\rightarrow 0$
and $p^2_Z\gg 1$. The renormalization scale $\mu^2$ enters implicitly by fixing the value for $\alpha(\mu^2)=g^2(\mu^2)/4\pi$.
In our calculations we set $Z(p^2_Z)=1$ which leads to $p^2_Z=\mu^2$. 
By using a \textit{perturbative} renormalization
scale of $\mu^2=5\times 10^4$ GeV$^2$ the condition $Z(\mu^2)=G(\mu^2)=1$ is valid, which is a special case of
the general form $Z(\mu^2)G^2(\mu^2)=1$ valid at all scales.
For the ghost boundary condition 
we used values in the range $G(0)^{-1}\in[1/16,1/22]$ which, below $N_f^{crit}$, result in the
decoupling-type solutions shown in figure~\ref{fig:1BCxG2_Results}.

\begin{figure}[!ht]
\center
\subfigure{\includegraphics[width=0.49\columnwidth]{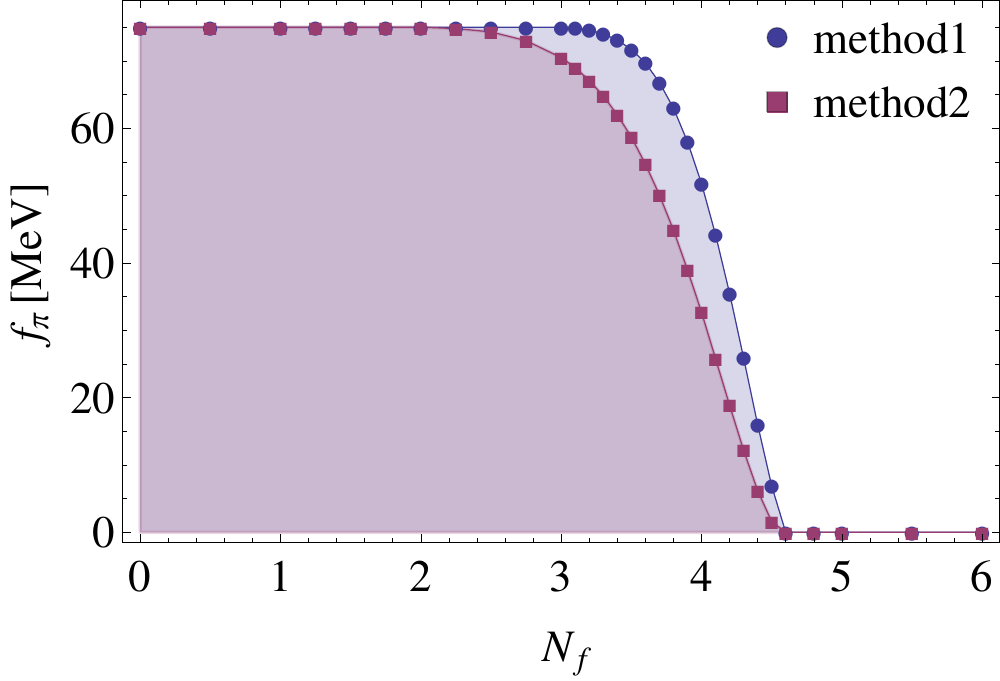}\label{fig:pion_decayconstant}}\hspace{0.1cm}
\subfigure{\includegraphics[width=0.49\columnwidth]{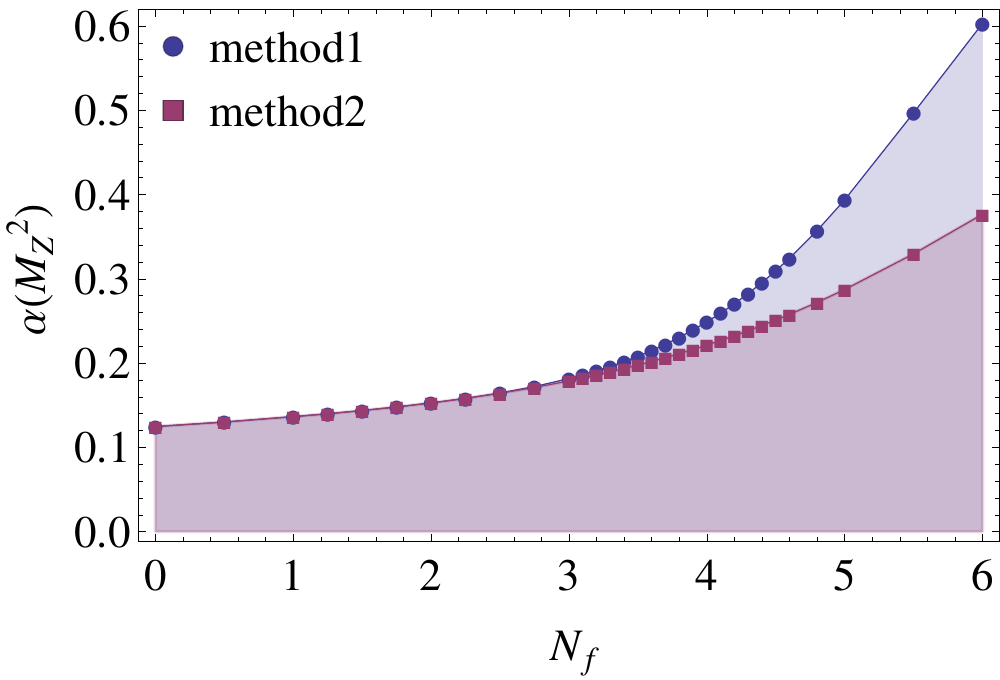}\label{fig:running_coupling_at_MZ}}
\subfigure{\includegraphics[width=0.49\columnwidth]{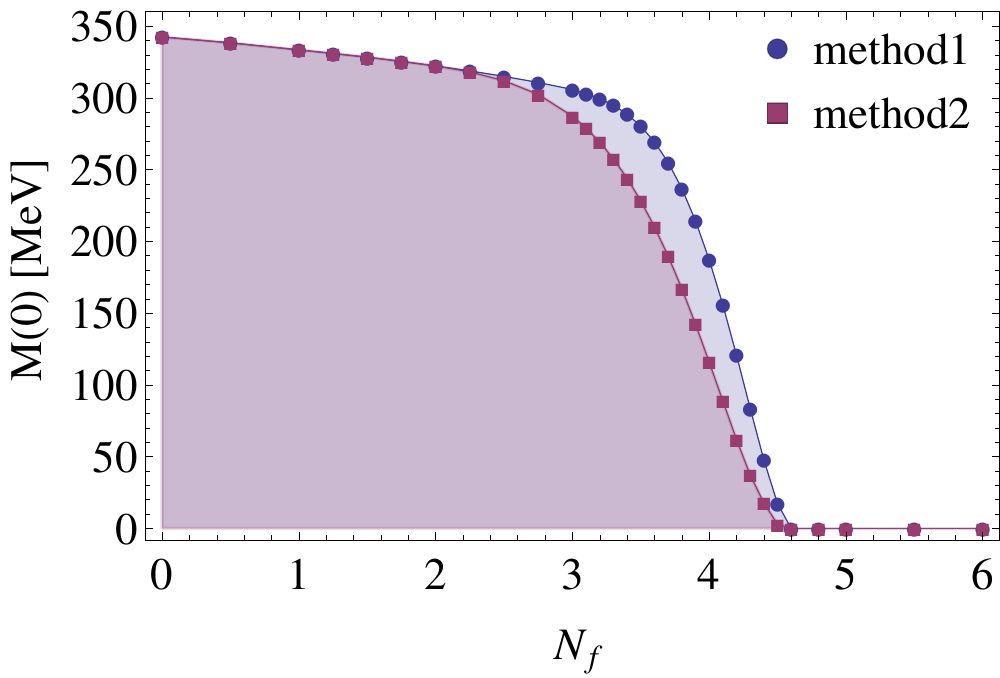}\label{fig:alpha_at_MZ}}\hspace{0.1cm}
\subfigure{\includegraphics[width=0.49\columnwidth]{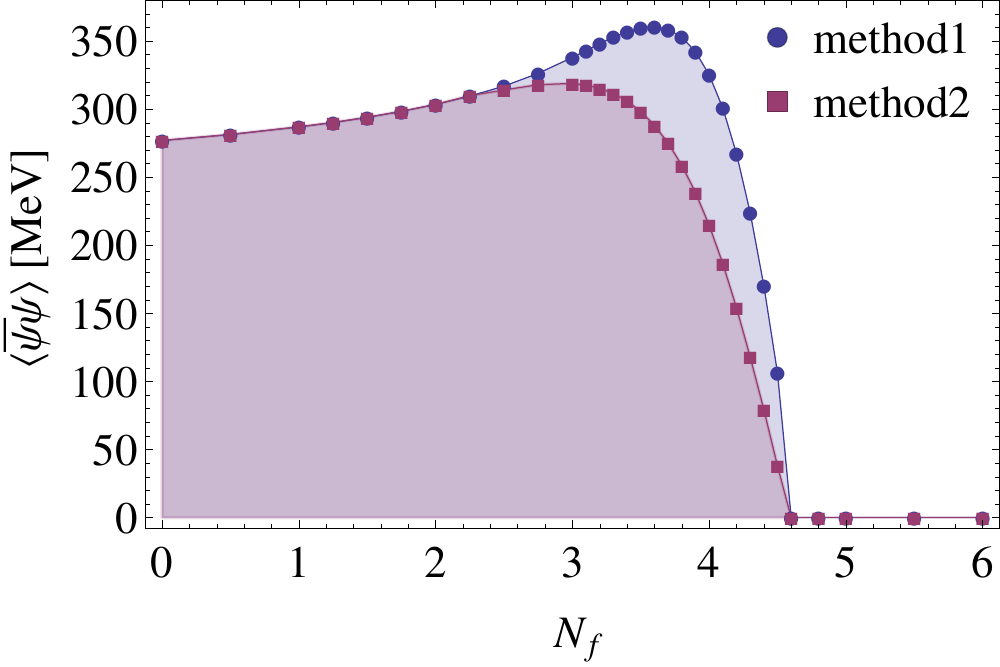}\label{fig:chiralcondensate_IRquarkmass}}
\caption{The pion decay constant $f_\pi$ fixed between $N_f\in[0,3]$ (method 1) and $N_f\in[0,2]$ (method 2), 
the running coupling $\alpha(M_Z^2)$ evaluated at the Z boson mass,
the infrared quark mass function $M(0)$ and the chiral condensate $\langle\bar\psi\psi\rangle$ for different flavours
using the scale fixing procedure detailed in appendix~\ref{sec:scale_fixing}.}
\label{fig:scale_fixing}
\end{figure}

In order to fix the system to an external (physical) scale a value for $\alpha(\mu^2)$ for a given $N_f$ 
needs to be specified. While this task is relatively easy if QCD is considered, an extrapolation
to a larger number of fermion flavours is highly non-trivial.
However, we want to stress a decisive point here. Even though the quantitative behaviour of the system
below the phase transition depends on the scale fixing, the specific value for $N_f^{crit}$ is 
independent of the scale fixing procedure to a high degree and solely depends on the 
truncations considered.
Hence, a pragmatic solution to fix the scale is to keep, {\it e.g.}, $f_\pi$ at a constant value 
within a reasonably small flavour number. It has been shown that the resulting behaviour of, {\it e.g.}, 
$M(p^2)$ is in good agreement with lattice results, {\it cf.} refs.~\cite{Fischer:2003rp,Fischer:2005en}. 
Our consideration is furthermore motivated by the fact that the 
approximation~\eqref{eq:PDC} underestimates $f_\pi$ within 10-20\% and
is not capable to reflect a correct flavour dependence of $f_\pi$.
From the specific values of the running coupling $\alpha(\mu^2)$ at small $N_f$ 
one can construct a flavour-dependent
coupling function which can then be extrapolated to larger $N_f$. 
In our calculations we keep $f_\pi=75\,MeV$ fixed between $N_f=0$ and $N_f=3$ and subsequently use 
standard Mathematica routines to extrapolate $\alpha(\mu^2)$. 
Results obtained from this procedure  are shown in figure~\ref{fig:scale_fixing}, where we also 
compare a fixing of $f_\pi$ between $N_f=0\rightarrow 2$.
As expected we observe quantitative differences below the phase transition. 
However, $N_f^{crit}$ is not affected by the scale fixing procedure.

\section{Influence of the Vertex Models}
\label{sec:vertex_impact}
In the following we investigate the influence of the models for the three-gluon and the
quark-gluon vertices used in our calculations.

\subsection{Relevance of the Quark-Gluon Vertex Model}
\label{sec:qgv_impact}
\begin{figure}[!ht]
\center
\subfigure{\includegraphics[height=0.30\columnwidth,width=0.5\columnwidth]{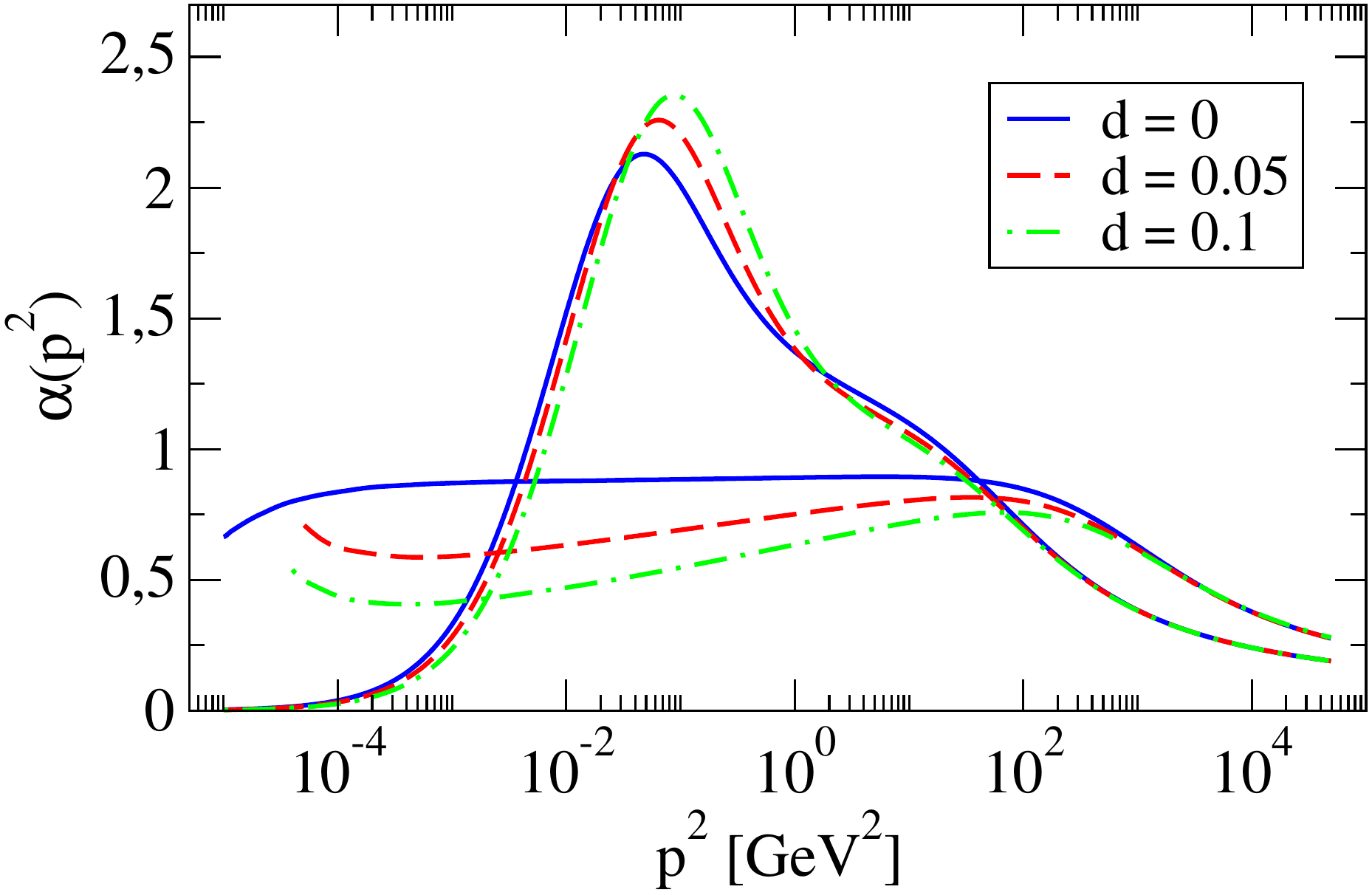}\label{fig:qgv_parameter_dependence1}}
\subfigure{\includegraphics[height=0.31\columnwidth,width=0.49\columnwidth]{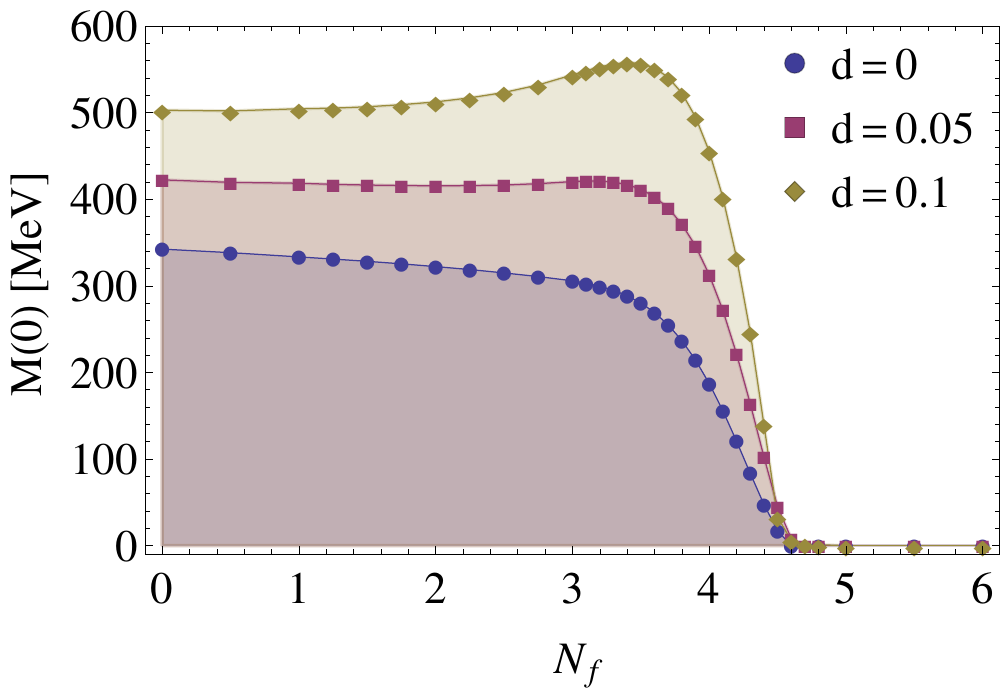}\label{fig:qgv_parameter_dependence2}} \\
\vspace{-0.3cm}
\subfigure{\includegraphics[height=0.30\columnwidth,width=0.5\columnwidth]{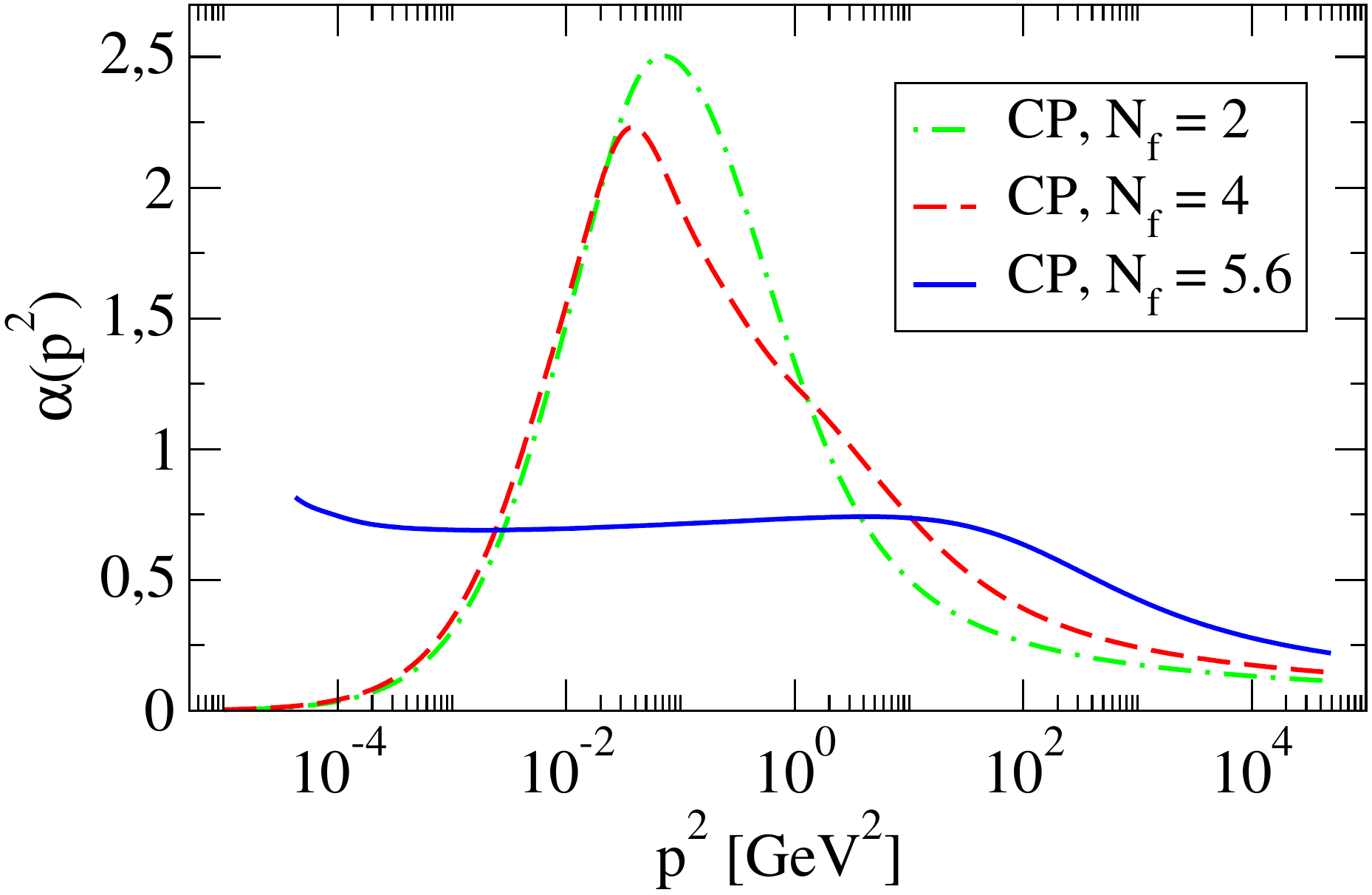}\label{fig:qgv_CP1}}
\subfigure{\includegraphics[height=0.31\columnwidth,width=0.49\columnwidth]{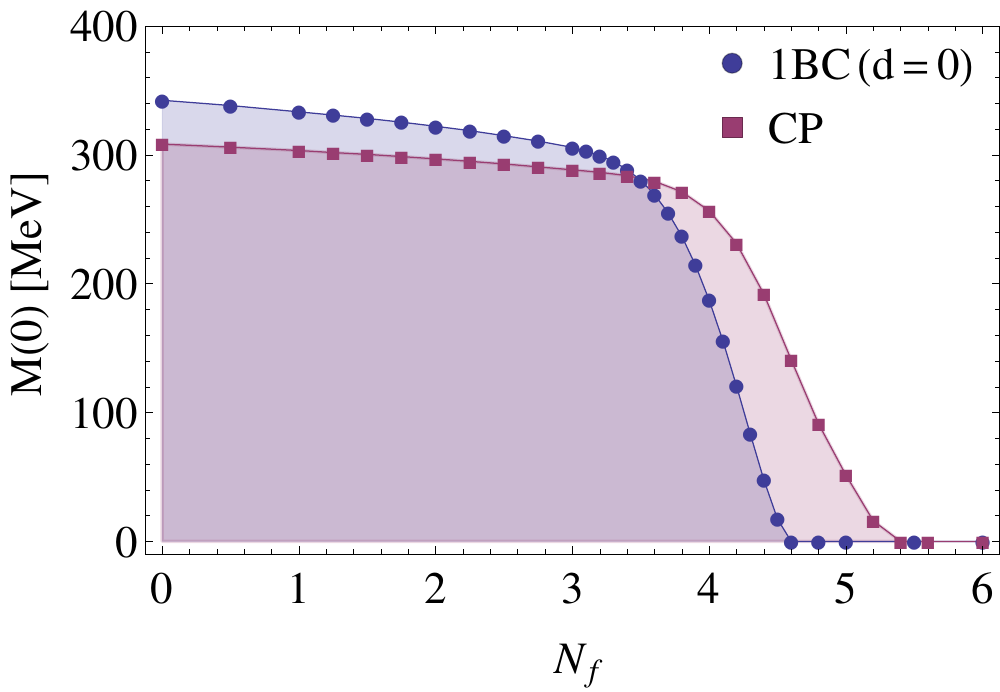}\label{fig:qgv_CP2}} \\
\vspace{-0.3cm}
\subfigure{\includegraphics[height=0.30\columnwidth,width=0.5\columnwidth]{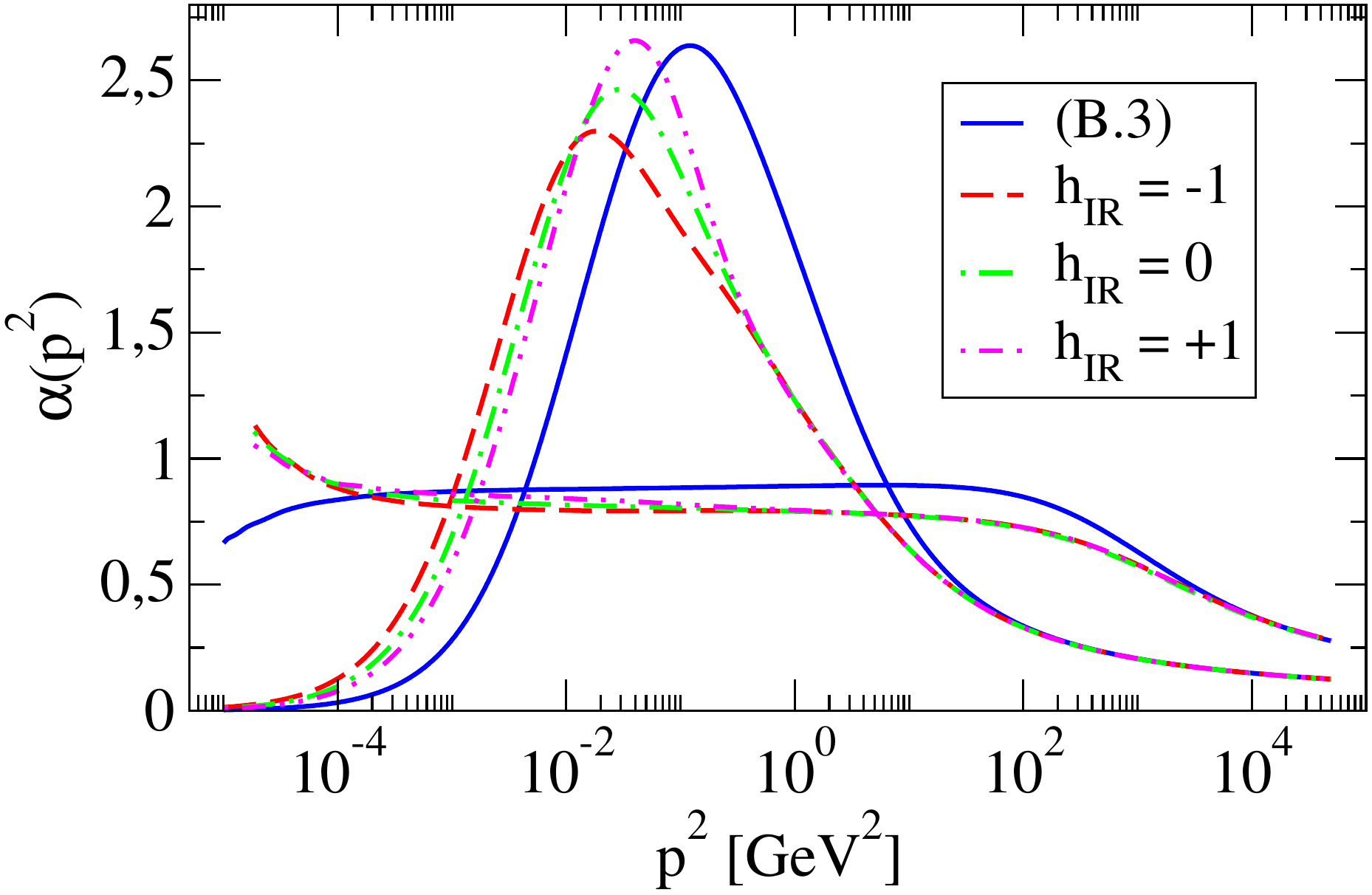}\label{fig:3gv_influence1}}
\subfigure{\includegraphics[height=0.31\columnwidth,width=0.49\columnwidth]{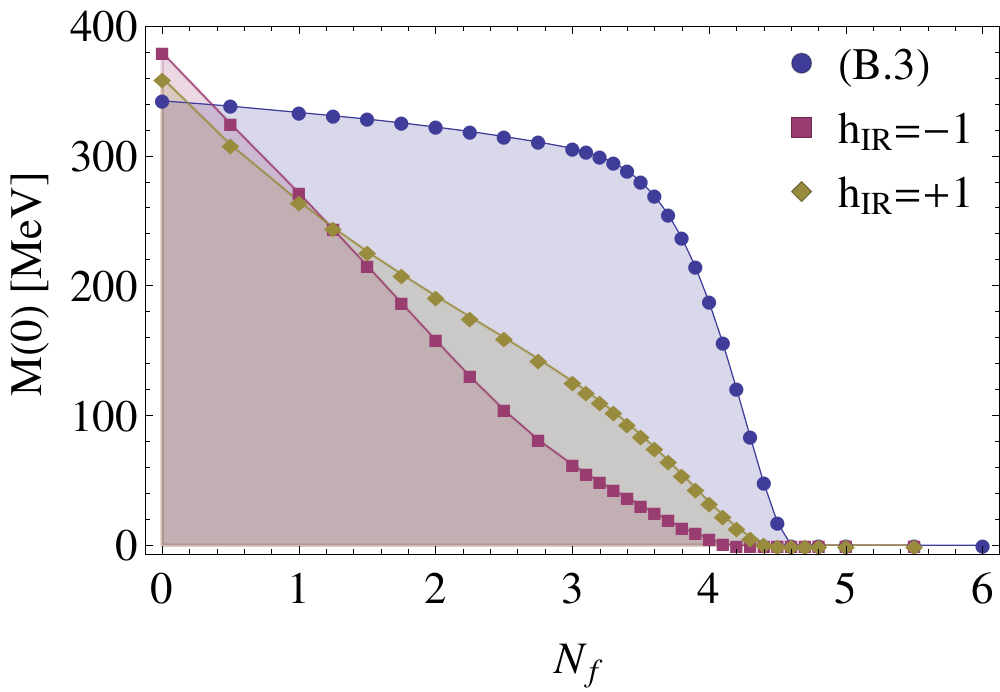}\label{fig:3gv_influence2}}
\vspace{-0.7cm}
\caption{
Upper Panel: The influence of the non-Abelian quark-gluon vertex parameter $d$.
Results for the running coupling 
below ($N_f=4$) and above ($N_f=5$) the phase transition (upper left). Whereas for larger values of $d$ 
the coupling becomes enhanced below $N_f^{crit}$ this effect turns over 
for $N_f\geq N_f^{crit}$ and a strong suppression is observed in this regime. 
Clearly, $N_f^{crit}$ is not affected by different parameter choices (upper right).
We note that negative values for $d$ tend to decrease $N_f^{crit}$ slightly since the effective 
interaction is to weak from the very beginning.
\newline
Middle Panel: Additional tensor structures tend to increase $N_f^{crit}$ as the calculation using a 
CP vertex reveals (middle right). The qualitative behaviour of the propagators and the 
running coupling is not changed though (middle left).
\newline
Lower Panel: The influence of the gauge boson vertex.
Although below $N_f^{crit}$ a dependence on the IR behaviour of the employed model is observed 
the system becomes remarkably IR independent above $N_f^{crit}$ (lower left). We use $N_f=2$ and $N_f=5$,
where solid lines correspond to eq.~\eqref{eq:3gv_Fischer}, dashed and dashed dotted lines 
represent results obtained for $h_{IR}\in\{-1,0,+1\}$ in eq.~\eqref{eq:3gv_Huber_IR}, respectively.
We kept $\Lambda_{3g}=1$ GeV fixed and did not fine-tune the model further.
In general different models for the three-gluon vertex tend to decrease $N_f^{crit}$ mildly (lower right).}
\label{fig:parameter_dependence}
\end{figure}

The quark-gluon vertex is the main ingredient in our calculation linking the Yang-Mills sector 
of the theory to the matter sector.
Thus, one expects a strong parameter dependence if this crucial object is replaced
by some model and indeed the rather low value of $N_f^{crit}$ can be traced back to this issue.
Our model presented in eq.~\eqref{eq:qgv_model} assumes a factorization into an Abelian part carrying the 
tensor structure and an effective non-Abelian interaction. 
A more general ansatz for the latter is given by \cite{Fischer:2003rp}
\begin{equation}
W^{\neg Abel}(p,q;k) = G^2(k^2) \:\tilde{Z}_3
\left(G(k^2) \: \tilde{Z}_3 \right)^{-2d-d/\delta}
 \Bigl(Z(k^2) \: {Z}_3\Bigr)^{-d}
\end{equation}
in case of eq.~\eqref{eq:vertex_nonabel} for the quark propagator and a symmetrized version
\begin{equation}
W^{\neg Abel}(q,k;p) = G(q^2)G(k^2) \:\tilde{Z}_3
\left(G(q^2)G(k^2) \: 
\tilde{Z}_3^2 \right)^{-d-d/(2\delta)}
\Bigl(Z(q^2)Z(k^2) \: {Z}_3^2\Bigr)^{-d/2}
\end{equation}
for the quark loop~\eqref{eq:vertex_quarkloop_nonabel},
where the effective interaction strength is controlled by the model parameter $d$.
When varying this parameter within reasonable bounds we find that $N_f^{crit}$ is 
not affected to a high degree, see figure~\ref{fig:qgv_parameter_dependence1} and
figure~\ref{fig:qgv_parameter_dependence2}.
Moreover, the stronger the effective quark-gluon interaction
gets below $N_f^{crit}$ the stronger is its suppression within the chirally symmetric phase. 
As different models for the  three-gluon vertex show no impact on the value of 
$N_f^{crit}$ (see the next subsection) we investigate the influence of 
additional tensor structures of the quark-gluon vertex\footnote{The 
beyond-tree-level tensor structures in the three-gluon vertex has been shown to be 
sub-leading \cite{Eichmann:2014xya}, {\it cf.} also ref.~\cite{Blum:2014gna}.
Thus, it is unlikely that they considerably influence $N_f^{crit}$.}.
In the figures~\ref{fig:qgv_CP1} and 
\ref{fig:qgv_CP2} we 
compare to results obtained from a Curtis-Pennington vertex
construction, {\it cf.} ref.~\cite{Fischer:2003rp} and references therein. 
In this case the additional structures clearly increase $N_f^{crit}$.
However, these vertex constructions are usually adapted from corresponding ansaetze in QED and might
reflect the correct behaviour of the quark-gluon vertex only partially.
Thus, a full calculation including all tensor structures seems to be inevitable in order to obtain
a more reliable value for $N_f^{crit}$.

\subsection{Relevance of the Three-Gluon Vertex Model}
\label{sec:3gv_impact}
For the three-gluon vertex we employ the tree-level structure and use two different models as well as variations of them 
in order to study the impact of non-perturbative contributions. 
We start with a model introduced in refs.~\cite{Fischer:2002eq, Fischer:2003rp}
\begin{equation}
\label{eq:3gv_Fischer}
 D^{3g}(p,q,k) = \frac{1}{Z_1}\frac{[G(q^2)G(k^2)]^{1-a/\delta-2a}}{[Z(q^2)Z(k^2)]^{1+a}} \ .
\end{equation}
Here, $Z_1$ is the renormalization constant for the three-gluon vertex and $a=3\delta$. 
This vertex model ensures by construction the correct logarithmic running of the gluon loop.
In ref.~\cite{Huber:2012kd} a Bose symmetric model was proposed which takes the form
\begin{equation}
\label{eq:3gv_Huber}
 D^{3g}(p,q,k) = \frac{1}{Z_1}\bigl[D_{IR}^{3g}(p,q,k) + D_{UV}^{3g}(p,q,k)\bigr]D_{UV}^{3g}(p,q,k) \ .
\end{equation}
The IR behaviour is described by
\begin{equation}
\label{eq:3gv_Huber_IR}
 D_{IR}^{3g}(p,q,k) = h_{IR}\,G(p^2+q^2+k^2)^3 \left[f^{3g}(p^2)f^{3g}(q^2)f^{3g}(k^2)\right]^4 \ ,
\end{equation}
with the auxiliary function $f^{3g}(p^2) = \Lambda_{3g}^2/(\Lambda_{3g}^2+p^2)$ and the IR parameter $h_{IR}$.
The UV part $D_{UV}^{3g}$ is detailed in ref.~\cite{Huber:2012kd}.
In figure~\ref{fig:3gv_influence1} and figure~\ref{fig:3gv_influence2} we show results obtained from the
different models.
In the broken phase, one observes large differences in the rate of change of the
quark mass $M(0)$, when the number of fermion flavours is enhanced. However, this does neither affect much
the location of the critical number of fermion flavours nor the qualitative behaviour of the theory in the 
symmetric phase\footnote{Furthermore, we note that with $\Lambda_{3g}$ an additional scale 
is introduced in case of Eq.~\eqref{eq:3gv_Huber_IR} which might change during the transition. 
On the other hand, the observed IR-independence 
provides substantial evidence that this would have only a small impact on $N_f^{crit}$.}.
Since the main point of our work is the latter, we conclude that our main results do not 
depend on the details of the vertex truncation.

\section{Removing spurious quadratic divergencies}
 \label{app:qDvgs_methods}

\begin{figure*}[!hb]
\center
\subfigure{\includegraphics[width=0.49\columnwidth]{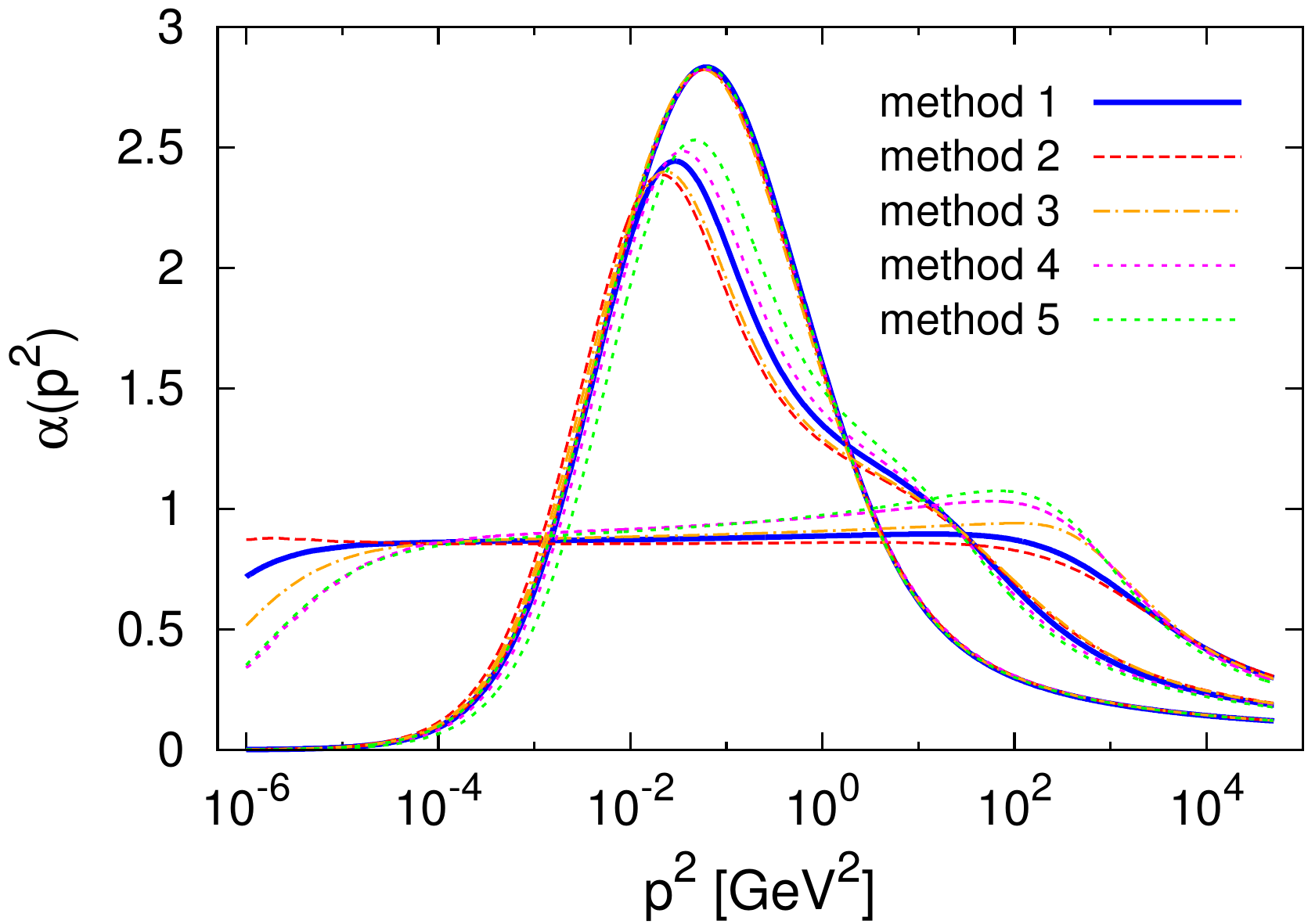}\label{fig:methods_coupling}}
\subfigure{\includegraphics[width=0.49\columnwidth]{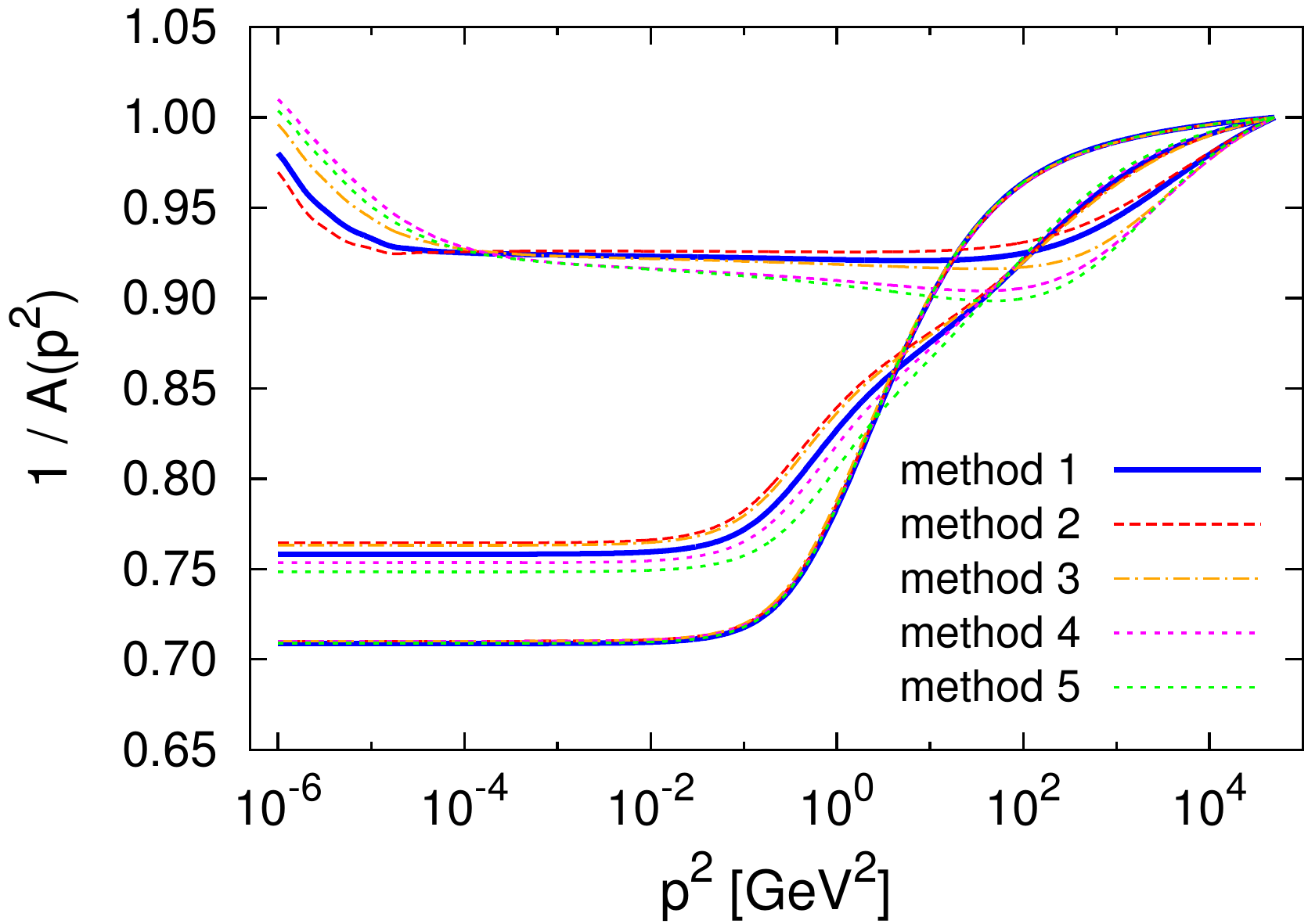}\label{fig:methods_A}}
\caption{A comparison of different subtraction methods for quadratic divergencies, where the corresponding
procedures are detailed in appendix~\ref{app:qDvgs_methods}.
The methods \textit{I} and \textit{II} correspond to a subtraction in the gluon loop including also
the simplification $Z_2A^{-1}(q^2\gg1)\approx 1$ and \textit{III} is the direct subtraction in the 
quark loop using a damping function. The numerical fit routine corresponds to method \textit{IV} 
(numerical subtraction only for the quark loop, the Yang-Mills system is treated conventionally)
and \textit{V} (numerical subtraction for all loops).
Results are presented for the running coupling $\alpha(p^2)$ and the quark wave-function 
renormalization $A^{-1}(p^2)$ for $N_f\in\{2,4,5\}$.
For small $N_f$ all methods are in excellent agreement. However, we observe small deviations
when the system comes close to the phase transition. The qualitative behaviour is not changed though.}
\label{fig:qDvgs_subtraction_results}
\end{figure*}

In general a truncated DSE system is plagued by spurious divergencies
appearing in the kernels of the loop integrals.
By contracting eq.~\eqref{eq:quarkloop_SE} with the generalized projection tensor
$\mathcal{P}_{\mu\nu}^{(\zeta)}(p) = \delta_{\mu\nu} - \zeta\, p_\mu p_\nu/p^2$
and setting $\zeta=4$ these contributions can be avoided.
However, such a procedure would disturb the IR behaviour of the system.
Based on a UV analysis a save way to get rid of these unwanted 
contributions is to modify the integral kernels by constructing appropriate compensation terms, {\it cf.}
refs.~\cite{%Fischer:2003zc,
Fischer:2003rp}.
For a moderate number of flavours the quark loop diagram is IR sub-leading 
such that a direct modification of the corresponding integral kernels is possible.
However, as soon as the system is within the chirally symmetric phase, {\it i.e.} 
if $M(p^2)\rightarrow 0$, the quark loop becomes IR enhanced and shows the same IR scaling behaviour
as the ghost loop. Hence, the method of subtracting quadratic divergencies directly
from the kernels of the quark loop fails if one wants to probe the chiral phase transition.
In the following we detail
several complementary methods which are able to eliminate these artificial contributions
in a safe way\footnote{An overview of different methods 
can also be found in ref.~\cite{Huber:2014tva}. Moreover, the novel subtraction scheme presented there might overcome
some of the obstacles inherent to present methods.},
where results are presented in figure~\ref{fig:qDvgs_subtraction_results}.

First, we describe a method to subtract quadratic divergencies from the 
gluon equation numerically \cite{Fischer:2005en}. 
The gluon DSE can schematically
be written as $Z^{-1}(x)=correct\;values + a / x$, where $x=p^2$ denotes the 
squared external momentum and
the fit parameter $a$ reflects the amount of the quadratic divergent 
contributions\footnote{We note that $a$ might contain (small) finite parts,
in particular if decoupling solutions are considered, 
which are subtracted as well as pointed out in ref.~\cite{Huber:2014tva}. 
Performing the fitting procedure in the UV should minimize these unwanted effects.}.
It depends on the loop kernels under consideration, {\it i.e.} on the applied
truncations/models and on the UV cutoff. Once this parameter is known the artificial
contributions can be subtracted, where in the following we briefly sketch the
procedure and refer to ref.~\cite{press:nr2007} for details.
In general a measured function $f$ is given by
\begin{equation}
 f(x) = \sum_{j=0}^{M-1}a_jX_j(x) = a_0X_0(x) + \ldots + a_{M-1}X_{M-1}(x) \ ,
\end{equation}
with fit parameters $a_j$ and fit functions $X_j(x)$.
For a given set of basis functions $X_k(x)$ one can define a merit function
\begin{equation}
 \chi^2 = \sum_{i=0}^{N-1}\biggl(f_i-\sum_{j=0}^{M-1}a_jX_j(x_i)\biggr)^2 \ ,
\end{equation}
which subsequently will be minimized in order to yield a set of optimized parameters $a_k$. 
Thus by setting $\partial \chi^2/\partial a_k = 0$ one obtains
 \begin{equation}
 \underbrace{\sum_{i=0}^{N-1}f_iX_k(x_i)}_{\beta_k} = 
 \sum_{j=0}^{M-1}a_j\underbrace{\sum_{i=0}^{N-1}X_j(x_i)X_k(x_i)}_{\alpha_{jk}} \ .
 \end{equation}
The corresponding matrix equation reads $\hat\alpha\,\boldsymbol a = \boldsymbol\beta$.
This procedure can now be applied to the gluon equation, which can be written as
$  Z^{-1}(x) = Z^{-1}(\sigma) + \Pi_Z(x) - \Pi_Z(\sigma) + a_1\left(x^{-1} - \sigma^{-1}\right)$.
The parameter $a_1$ is obtained by fitting the gluon dressing function 
in a region where its behaviour can be derived 
analytically either from an infrared analysis or from perturbation theory.
For large momenta $x\geq x_{UV}$ the gluon dressing function takes the perturbative form
\begin{equation}
 Z^{-1}(x) = \frac{a_0}{Z(x_{UV})}\left[\omega\log\left(\frac{x}{x_{UV}}\right)+1\right]^{-\gamma} + a_1\left(x^{-1} - \sigma^{-1}\right)
\end{equation}
and the fitting procedure is performed using 
$X_0(x_i) = Z^{-1}(x_{UV})\left[\omega\log\left(x_i/x_{UV}\right)+1\right]^{-\gamma}$
and $X_1(x_i) = x_i^{-1} - \sigma^{-1}$
as well as $f_i = Z^{-1}(x_i) = Z^{-1}(\sigma) + \Pi_Z(x_i) - \Pi_Z(\sigma)$ with $\omega=(11N_c-2N_f)\alpha(\mu^2)/(12\pi)$.
The advantage of a UV fitting is that the IR behaviour 
of the gluon propagator is unaffected from the subtraction to a high degree.
This is particularly important when investigating the chiral phase transition
where the IR behaviour changes drastically.
The inversion of $\hat\alpha$ is simple and the fit parameters are given by\footnote{The numerical subtraction requires only $a_1$. 
Thus, $a_0$ may act as an additional control parameter.}
$a_0 = \left(\alpha_{11}\beta_0-\alpha_{01}\beta_1\right)/det\,\hat\alpha$
and $a_1 = \left(\alpha_{00}\beta_1-\alpha_{01}\beta_0\right)/det\,\hat\alpha$,
where $det\,\hat\alpha = \alpha_{00}\alpha_{11}-\alpha_{01}^2$ and $\alpha_{01}=\alpha_{10}$.

In order to eliminate quadratically divergent terms analytically a UV analysis has to be performed,
{\it cf. e.g.} ref.~\cite{Fischer:2003rp}.
The self-energy contribution from the quark loop is given in eq.~\eqref{eq:quarkloop_SE_contracted}. 
In the far UV the approximation $f(k^2)\approx f(q^2)$ 
for the dressing functions $f\in\{A,B,G,Z\}$ is valid. Furthermore, the 
mass term in the denominator of the auxiliary function $\sigma_v$
can be neglected. Thus, the angular integration can be performed leaving the following simplified 
expression for the quark loop self-energy contribution
\begin{equation}
 \Pi^{UV}_{quark}(p^2) = -\frac{g^2N_f}{16\pi^2}\int dq^2\,\frac{G(q^2)^{2-2d-d/\delta}}{Z(q^2)^d}\frac{Z_2}{A(q^2)}
\left(\frac{-2}{3q^2}+\frac{4-\zeta}{3p^2}+\frac{2(4-\zeta)}{3p^2q^2}M^2(q^2)\right) \ .
\label{eq:quadDvgs_quarkloop}
\end{equation}
The anomalous dimensions for the Yang-Mills propagators fulfill the relation 
$1+\gamma+2\delta=0$ and thus we obtain
$G(q^2)^{-2d-d/\delta} Z(q^2)^{-d} = \left[\omega \log\left(p^2/\mu^2\right)+1\right]^{-d(1+\gamma+2\delta)} = 1$,
where we used the renormalization condition $Z(\mu^2)=G(\mu^2)=1$ which is
valid for a perturbative renormalization scale $\mu^2$.
Hence, the effective non-Abelian interactions for the quark loop
in the far UV are governed by the factor $G(q^2)^2$ and are independent of the parameter $d$.
One can see that from the last term in eq.~\eqref{eq:quadDvgs_quarkloop} an unphysical longitudinal contribution origins. 
Although this contribution is not quadratically divergent it is an artifact of the truncation 
and must be subtracted in order to render the gluon propagator transverse in the UV.
The DSE for the gluon propagator in the far UV finally reads
\begin{eqnarray}
 \frac{1}{Z(p^2)} = Z_3 & + & \frac{g^2N_c}{48\pi^2}\int \frac{dq^2}{p^2}\biggl[
       \left(\frac{4-\zeta}{4}+\frac{\zeta-2}{4}\frac{p^2}{q^2}\right)G(q^2)^2 \nonumber \\
 & + & \left(\frac{-6(4-\zeta)}{4}-\frac{\zeta+24}{4}\frac{p^2}{q^2}+\frac{7}{8}\frac{p^4}{q^4}\right)\frac{G(q^2)^{-4-12\delta}}{Z(q^2)^{6\delta}} \\
 & - & \left(\frac{-2p^2}{q^2}+\frac{4(4-\zeta)}{4}\right)G(q^2)^2\frac{N_f}{Nc}Z_2A^{-1}(q^2) \nonumber
\biggr] \ ,
\end{eqnarray}
where the three terms correspond to the ghost, gluon and quark loop contribution, respectively, 
{\it cf. e.g.} refs.~\cite{Fischer:2003rp}.
Hence, the quadratically divergent terms read
\begin{eqnarray}
\label{eq:gluon_eq_UV_quadDvgs}
 \frac{1}{Z(p^2)} 
%  & = & Z_3 + \frac{g^2N_c}{48\pi^2}\int \frac{dq^2}{p^2}G(q^2)^2\biggl[
%  \frac{4-\zeta}{4} - \frac{6(4-\zeta)}{4} - \frac{4(4-\zeta)}{4}\frac{N_f}{Nc}Z_2A^{-1}(q^2)\biggr] \nonumber \\
 & = & Z_3 + \frac{g^2N_c}{48\pi^2}\int \frac{dq^2}{p^2}G(q^2)^2\biggl[
 -\frac{5}{4} - \frac{N_f}{Nc}Z_2A^{-1}(q^2)+\ldots\biggr](4-\zeta) \ ,
\end{eqnarray}
where we used that $G(q^2)^{-4-12\delta}/Z(q^2)^{6\delta} = G(q^2)^2$ in the UV.
Thus, in order to compensate the divergent terms of all three loops
one can modify the kernel of the gluon loop by adding the following additional terms
\begin{equation}
 Q(p^2,q^2,k^2) \rightarrow Q(p^2,q^2,k^2) + \left(\frac{5}{4}+\frac{N_f}{Nc}Z_2A^{-1}(q^2)\right)(4-\zeta) \ .
\end{equation}
Since the gluon loop is sub-leading in the IR regime\footnote{Similar to the method introduced in the following
it acts as a natural damping function.} the new compensation terms do not influence the 
IR behaviour of the system.
Furthermore, by using $A(q^2)\stackrel{q^2\rightarrow \infty}{\rightarrow} Z_2$ 
we can additionally simplify this expression by setting 
$Z_2A^{-1}(q^2)\approx 1$ since the integral should be dominated by the external momentum scale. 

The idea of the last method we want to present is to subtract the unwanted contributions directly 
in the corresponding kernels.
In order not to influence the IR and mid-momentum regime the compensation terms have to be suppressed 
in these regions by damping functions.
Based on the UV analysis of ref.~\cite{Fischer:2003rp} the kernel $U$ given in eq.~\eqref{eq:U} is modified as follows
\begin{equation}
 U(p^2,q^2,k^2) \rightarrow U(p^2,q^2,k^2) - (4-\zeta)\frac{q^2+k^2}{6p^2}\,f_{UV}(q^2) \ ,
\end{equation}
where the damping function $f_{UV}(q^2) = \tanh(q^2\,\xi^2)$ is taken from ref.~\cite{Huber:2012kd}
and we use $\xi\approx0.5\,GeV$ for the damping parameter.
We note that if $\xi$ is chosen too small the IR behaviour of the quark loop 
is influenced and $N_f^{crit}$ is higher in this
case. This however interferes with the results obtained from the other methods. 
Hence, a direct subtraction in the quark loop without a damping function is problematic for $N_f\gtrsim N_f^{crit}$. 
For the ghost and gluon loop kernels this procedure is analogous and explained in detail
in ref.~\cite{Huber:2012kd}.

%%%%%%%%%%%%%%%%%%%%%%%%%%%%%%%%%%%%%%%%%%%%%%%%%%%%%%%%%%%%%%%%%%%%%%%%%%%%%%%%%%%%%%%%%%%%%%%%%%%%%%%%%%%%%%%%%


\begin{thebibliography}{99}
%\cite{Farhi:1980xs}
\bibitem{Farhi:1980xs}
  E.~Farhi and L.~Susskind,
  %``Technicolor,''
  Phys.\ Rept.\  {\bf 74} (1981) 277;
  %%CITATION = PRPLC,74,277;%%
  %911 citations counted in INSPIRE as of 20 Nov 2013
%\cite{Hill:2002ap}
% \bibitem{Hill:2002ap}
  C.~T.~Hill and E.~H.~Simmons,
  %``Strong dynamics and electroweak symmetry breaking,''
  Phys.\ Rept.\  {\bf 381} (2003) 235
   [Erratum-ibid.\  {\bf 390} (2004) 553]
  [hep-ph/0203079].
  %%CITATION = HEP-PH/0203079;%%
  %704 citations counted in INSPIRE as of 20 Nov 2013

%\cite{Weinberg:1979bn}
\bibitem{Weinberg:1979bn}
  S.~Weinberg,
  %``Implications of Dynamical Symmetry Breaking: An Addendum,''
  Phys.\ Rev.\ D {\bf 19} (1979) 1277;
  %%CITATION = PHRVA,D19,1277;%%
  %1552 citations counted in INSPIRE as of 20 Nov 2013
%\cite{Susskind:1978ms}
% \bibitem{Susskind:1978ms}
  L.~Susskind,
  %``Dynamics of Spontaneous Symmetry Breaking in the Weinberg-Salam Theory,''
  Phys.\ Rev.\ D {\bf 20} (1979) 2619;
  %%CITATION = PHRVA,D20,2619;%%
  %1957 citations counted in INSPIRE as of 20 Nov 2013
%\cite{}
% \bibitem{Weinberg:1976}
  S.~Weinberg,
  %``Implications of Dynamical Symmetry Breaking,''
  Phys.\ Rev.\ D {\bf 13} (1976) 974.
  %%CITATION = PHRVA,D13,974;%%
  %1213 citations counted in INSPIRE as of 20 Nov 2013

%\cite{Dimopoulos:1979es}
\bibitem{Dimopoulos:1979es}
  S.~Dimopoulos and L.~Susskind,
  %``Mass Without Scalars,''
  Nucl.\ Phys.\ B {\bf 155} (1979) 237.
  %%CITATION = NUPHA,B155,237;%%
  %973 citations counted in INSPIRE as of 20 Nov 2013

%\cite{Eichten:1979ah}
\bibitem{Eichten:1979ah}
  E.~Eichten and K.~D.~Lane,
  %``Dynamical Breaking of Weak Interaction Symmetries,''
  Phys.\ Lett.\ B {\bf 90} (1980) 125.
  %%CITATION = PHLTA,B90,125;%%
  %989 citations counted in INSPIRE as of 20 Nov 2013

%\cite{Glashow:1970gm}
\bibitem{Glashow:1970gm}
  S.~L.~Glashow, J.~Iliopoulos and L.~Maiani,
  %``Weak Interactions with Lepton-Hadron Symmetry,''
  Phys.\ Rev.\ D {\bf 2} (1970) 1285.
  %%CITATION = PHRVA,D2,1285;%%
  %4556 citations counted in INSPIRE as of 20 Nov 2013

%\cite{Holdom:1984sk}
\bibitem{Holdom:1984sk}
  B.~Holdom,
  %``Techniodor,''
  Phys.\ Lett.\ B {\bf 150} (1985) 301;
  %%CITATION = PHLTA,B150,301;%%
  %489 citations counted in INSPIRE as of 20 Nov 2013
%\cite{Akiba:1985rr}
% \bibitem{Akiba:1985rr}
  T.~Akiba and T.~Yanagida,
  %``Hierarchic Chiral Condensate,''
  Phys.\ Lett.\ B {\bf 169} (1986) 432.
  %%CITATION = PHLTA,B169,432;%%
  %283 citations counted in INSPIRE as of 20 Nov 2013

%\cite{Yamawaki:1985zg}
\bibitem{Yamawaki:1985zg}
  K.~Yamawaki, M.~Bando and K.~-i.~Matumoto,
  %``Scale Invariant Technicolor Model and a Technidilaton,''
  Phys.\ Rev.\ Lett.\  {\bf 56} (1986) 1335.
  %%CITATION = PRLTA,56,1335;%%
  %650 citations counted in INSPIRE as of 20 Nov 2013
%\cite{Bando:1986bg}
% \bibitem{Bando:1986bg}
  M.~Bando, K.~-i.~Matumoto and K.~Yamawaki,
  %``Technidilaton,''
  Phys.\ Lett.\ B {\bf 178} (1986) 308;
  %%CITATION = PHLTA,B178,308;%%
  %121 citations counted in INSPIRE as of 20 Nov 2013
%\cite{Bando:1987we}
% \bibitem{Bando:1987we}
  M.~Bando, T.~Morozumi, H.~So and K.~Yamawaki,
  %``Discriminating Technicolor Theories Through Flavor Changing Neutral Currents: Walking Or Standing Coupling Constants?,''
  Phys.\ Rev.\ Lett.\  {\bf 59} (1987) 389.
  %%CITATION = PRLTA,59,389;%%
  %120 citations counted in INSPIRE as of 20 Nov 2013

%\cite{Appelquist:1986an}
\bibitem{Appelquist:1986an}
  T.~W.~Appelquist, D.~Karabali and L.~C.~R.~Wijewardhana,
  %``Chiral Hierarchies and the Flavor Changing Neutral Current Problem in Technicolor,''
  Phys.\ Rev.\ Lett.\  {\bf 57} (1986) 957;
  %%CITATION = PRLTA,57,957;%%
  %647 citations counted in INSPIRE as of 20 Nov 2013
%\cite{}
% \bibitem{}
  T.~Appelquist and L.~C.~R.~Wijewardhana.
  %``Chiral Hierarchies from Slowly Running Couplings in Technicolor Theories,''
  Phys.\ Rev.\ D {\bf 36} (1987) 568.
  %%CITATION = PHRVA,D36,568;%%
  %383 citations counted in INSPIRE as of 20 Nov 2013

%\cite{Lane:2002wv}
\bibitem{Lane:2002wv}
  K.~Lane,
  %``Two lectures on technicolor,''
  hep-ph/0202255.
  %%CITATION = HEP-PH/0202255;%%
  %153 citations counted in INSPIRE as of 22 Nov 2013

%\cite{Sannino:2009za}
\bibitem{Sannino:2009za}
  F.~Sannino,
  %``Conformal Dynamics for TeV Physics and Cosmology,''
  Acta Phys.\ Polon.\ B {\bf 40} (2009) 3533
  [arXiv:0911.0931 [hep-ph]].
  %%CITATION = ARXIV:0911.0931;%%
  %156 citations counted in INSPIRE as of 22 Nov 2013

%\cite{Lane:1991qh}
\bibitem{Lane:1991qh}
  K.~D.~Lane and M.~V.~Ramana,
  %``Walking technicolor signatures at hadron colliders,''
  Phys.\ Rev.\ D {\bf 44} (1991) 2678.
  %%CITATION = PHRVA,D44,2678;%%
  %173 citations counted in INSPIRE as of 20 Nov 2013

%\cite{Appelquist:1997fp}
\bibitem{Appelquist:1997fp}
  T.~Appelquist, J.~Terning and L.~C.~R.~Wijewardhana,
  %``Postmodern technicolor,''
  Phys.\ Rev.\ Lett.\  {\bf 79} (1997) 2767
  [hep-ph/9706238];
  %%CITATION = HEP-PH/9706238;%%
  %69 citations counted in INSPIRE as of 20 Nov 2013

%\cite{}
\bibitem{Caswell:1974}
  W.~E.~Caswell,
  %``Asymptotic Behavior of Nonabelian Gauge Theories to Two Loop Order,''
  Phys.\ Rev.\ Lett.\  {\bf 33} (1974) 244.
  %%CITATION = PRLTA,33,244;%%
  %663 citations counted in INSPIRE as of 20 Nov 2013

%\cite{}
\bibitem{Banks:1982}
  T.~Banks and A.~Zaks,
  %``On the Phase Structure of Vector-Like Gauge Theories with Massless Fermions,''
  Nucl.\ Phys.\ B {\bf 196} (1982) 189.
  %%CITATION = NUPHA,B196,189;%%
  %618 citations counted in INSPIRE as of 20 Nov 2013

%\cite{Appelquist:1996dq}
\bibitem{Appelquist:1996dq}
  T.~Appelquist, J.~Terning and L.~C.~R.~Wijewardhana,
  %``The Zero temperature chiral phase transition in SU(N) gauge theories,''
  Phys.\ Rev.\ Lett.\  {\bf 77} (1996) 1214
  [hep-ph/9602385];
  %%CITATION = HEP-PH/9602385;%%
  %223 citations counted in INSPIRE as of 22 Nov 2013
%\cite{Appelquist:1998rb}
% \bibitem{Appelquist:1998rb}
  T.~Appelquist, A.~Ratnaweera, J.~Terning and L.~C.~R.~Wijewardhana,
  %``The Phase structure of an SU(N) gauge theory with N(f) flavours,''
  Phys.\ Rev.\ D {\bf 58} (1998) 105017
  [hep-ph/9806472].
  %%CITATION = HEP-PH/9806472;%%
  %137 citations counted in INSPIRE as of 22 Nov 2013

%\cite{Gies:2005as}
\bibitem{Gies:2005as}
  H.~Gies and J.~Jaeckel,
  %``Chiral phase structure of QCD with many flavours,''
  Eur.\ Phys.\ J.\ C {\bf 46} (2006) 433
  [hep-ph/0507171].
  %%CITATION = HEP-PH/0507171;%%
  %74 citations counted in INSPIRE as of 22 Nov 2013


%\cite{Appelquist:2007hu}
\bibitem{Appelquist:2007hu}
  T.~Appelquist, G.~T.~Fleming and E.~T.~Neil,
  %``Lattice study of the conformal window in QCD-like theories,''
  Phys.\ Rev.\ Lett.\  {\bf 100} (2008) 171607
   [Erratum-ibid.\  {\bf 102} (2009) 149902]
  [arXiv:0712.0609 [hep-ph]].
  %%CITATION = ARXIV:0712.0609;%%
  %204 citations counted in INSPIRE as of 22 Nov 2013

%\cite{Appelquist:2009ty}
\bibitem{Appelquist:2009ty}
  T.~Appelquist, G.~T.~Fleming and E.~T.~Neil,
  %``Lattice Study of Conformal Behavior in SU(3) Yang-Mills Theories,''
  Phys.\ Rev.\ D {\bf 79} (2009) 076010
  [arXiv:0901.3766 [hep-ph]];
  %%CITATION = ARXIV:0901.3766;%%
  %166 citations counted in INSPIRE as of 22 Nov 2013
%\cite{Deuzeman:2009mh}
% \bibitem{Deuzeman:2009mh}
  A.~Deuzeman, M.~P.~Lombardo and E.~Pallante,
  %``Evidence for a conformal phase in SU(N) gauge theories,''
  Phys.\ Rev.\ D {\bf 82} (2010) 074503
  [arXiv:0904.4662 [hep-ph]].
  %%CITATION = ARXIV:0904.4662;%%
  %133 citations counted in INSPIRE as of 22 Nov 2013
%\cite{Appelquist:2011dp}
% \bibitem{Appelquist:2011dp}
  T.~Appelquist, G.~T.~Fleming, M.~F.~Lin, E.~T.~Neil and D.~A.~Schaich,
  %``Lattice Simulations and Infrared Conformality,''
  Phys.\ Rev.\ D {\bf 84} (2011) 054501
  [arXiv:1106.2148 [hep-lat]].
  %%CITATION = ARXIV:1106.2148;%%
  %54 citations counted in INSPIRE as of 22 Nov 2013
%\cite{Hasenfratz:2011xn}
% \bibitem{Hasenfratz:2011xn}
  A.~Hasenfratz,
  %``Infrared fixed point of the 12-fermion SU(3) gauge model based on 2-lattice MCRG matching,''
  Phys.\ Rev.\ Lett.\  {\bf 108} (2012) 061601
  [arXiv:1106.5293 [hep-lat]].
  %%CITATION = ARXIV:1106.5293;%%
  %38 citations counted in INSPIRE as of 22 Nov 2013

%\cite{Fodor:2012et}
\bibitem{Fodor:2012et}
  Z.~Fodor, K.~Holland, J.~Kuti, D.~Nogradi, C.~Schroeder and C.~H.~Wong,
  %``Conformal finite size scaling of twelve fermion flavours,''
  PoS LATTICE {\bf 2012} (2012) 279
  [arXiv:1211.4238 [hep-lat]].
  %%CITATION = ARXIV:1211.4238;%%
  %12 citations counted in INSPIRE as of 22 Nov 2013

%\cite{Braun:2009ns}
\bibitem{Braun:2009ns}
  J.~Braun and H.~Gies,
  %``Scaling laws near the conformal window of many-flavor QCD,''
  JHEP {\bf 1005} (2010) 060
  [arXiv:0912.4168 [hep-ph]].
  %%CITATION = ARXIV:0912.4168;%%
  %33 citations counted in INSPIRE as of 16 May 2014

%\cite{Braun:2010qs}
\bibitem{Braun:2010qs}
  J.~Braun, C.~S.~Fischer and H.~Gies,
  %``Beyond Miransky Scaling,''
  Phys.\ Rev.\ D {\bf 84} (2011) 034045
  [arXiv:1012.4279 [hep-ph]].
  %%CITATION = ARXIV:1012.4279;%%
  %31 citations counted in INSPIRE as of 16 May 2014



%\cite{Alkofer:2000wg}
\bibitem{Alkofer:2000wg}
  R.~Alkofer and L.~von Smekal,
  %``The infrared behaviour of QCD Green's functions: Confinement, dynamical
  %symmetry breaking, and hadrons as relativistic bound states,''
  Phys.\ Rept.\  {\bf 353} (2001) 281
  [arXiv:hep-ph/0007355];
  %%CITATION = PRPLC,353,281;%%
  %\cite{Fischer:2006ub}
% \bibitem{Fischer:2006ub}
  C.~S.~Fischer,
  %``Infrared properties of QCD from Dyson-Schwinger equations,''
  J.\ Phys.\ G {\bf 32} (2006) R253
  [hep-ph/0605173].
  %%CITATION = HEP-PH/0605173;%%
  %267 citations counted in INSPIRE as of 20 Nov 2013

%  \cite{Maris:2003vk}
\bibitem{Maris:2003vk}
P.~Maris and C.D.~Roberts,
%``Dyson-Schwinger equations: A tool for hadron physics,''
Int.\ J.Mod.\ Phys.\ E{\bf 12} (2003) 297
[arXiv:nucl-th/0301049].
%%CITATION = NUCL-TH 0301049;%%

%\cite{Alkofer:2004it}
\bibitem{Alkofer:2004it}
  R.~Alkofer, C.~S.~Fischer and F.~J.~Llanes-Estrada,
  %``Vertex functions and infrared fixed point in Landau gauge SU(N) Yang-Mills theory,''
  Phys.\ Lett.\ B {\bf 611} (2005) 279
   [Erratum-ibid.\  {\bf 670} (2009) 460]
  [hep-th/0412330];
  %%CITATION = HEP-TH/0412330;%%
%\cite{Huber:2007kc}
% \bibitem{Huber:2007kc}
  M.~Q.~Huber, R.~Alkofer, C.~S.~Fischer and K.~Schwenzer,
  %``The Infrared behaviour of Landau gauge Yang-Mills theory in d=2, d=3 and d=4 dimensions,''
  Phys.\ Lett.\ B {\bf 659} (2008) 434
  [arXiv:0705.3809 [hep-ph]];
  %%CITATION = ARXIV:0705.3809;%%
  %58 citations counted in INSPIRE as of 20 Nov 2013
%\cite{Fischer:2009tn}
%\bibitem{Fischer:2009tn}
  C.~S.~Fischer and J.~M.~Pawlowski,
  %``Uniqueness of infrared asymptotics in Landau gauge Yang-Mills theory II,''
  Phys.\ Rev.\ D {\bf 80} (2009) 025023
  [arXiv:0903.2193 [hep-th]].
  %%CITATION = ARXIV:0903.2193;%%
  %72 citations counted in INSPIRE as of 01 May 2014

%\cite{Bashir:2013zha}
\bibitem{Bashir:2013zha}
  A.~Bashir, A.~Raya and J.~Rodriguez-Quintero,
  %``QCD: Restoration of Chiral Symmetry and Deconfinement for Large N_f,''
  Phys.\ Rev.\ D {\bf 88} (2013) 054003
  [arXiv:1302.5829 [hep-ph]].
  %%CITATION = ARXIV:1302.5829;%%
  %5 citations counted in INSPIRE as of 15 Apr 2014

%\cite{Fischer:2003rp}
 \bibitem{Fischer:2003rp}
  C.~S.~Fischer and R.~Alkofer,
  %``Nonperturbative propagators, running coupling and dynamical quark mass of Landau gauge QCD,''
  Phys.\ Rev.\ D {\bf 67} (2003) 094020
  [hep-ph/0301094];
  %%CITATION = HEP-PH/0301094;%%
%\cite{Fischer:2003zc}
% \bibitem{Fischer:2003zc}
  C.~S.~Fischer,
  %``Nonperturbative propagators, running coupling and dynamical mass generation in ghost - anti-ghost symmetric gauges in QCD,''
  PhD thesis [hep-ph/0304233].
  %%CITATION = HEP-PH/0304233;%%
  %35 citations counted in INSPIRE as of 28 Oct 2013

%\cite{Mader:2013ru}
\bibitem{Mader:2013ru}
  V.~Mader and R.~Alkofer,
  %``Including 4-gluon interactions into Dyson-Schwinger studies,''
  PoS ConfinementX {\bf } (2012) 063
  [arXiv:1301.7498 [hep-th]].
  %%CITATION = ARXIV:1301.7498;%%
  %2 citations counted in INSPIRE as of 28 Oct 2013

%\cite{Huber:2012kd}
\bibitem{Huber:2012kd}
  M.~Q.~Huber and L.~von Smekal,
  %``On the influence of three-point functions on the propagators of Landau gauge Yang-Mills theory,''
  JHEP {\bf 1304} (2013) 149
  [arXiv:1211.6092 [hep-th]];
  %%CITATION = ARXIV:1211.6092;%%
  %25 citations counted in INSPIRE as of 27 May 2014
%\cite{Huber:2013xb}
% \bibitem{Huber:2013xb}
 M.~Q.~Huber and L.~von Smekal,
  %``Going beyond the propagators of Landau gauge Yang-Mills theory,''
  PoS CONFINEMENTX {\bf } (2013) 062
  [arXiv:1301.3080 [hep-th]].
  %%CITATION = ARXIV:1301.3080;%%  

\bibitem{Cucchieri:2004sq}
  A.~Cucchieri, T.~Mendes and A.~Mihara,
  JHEP {\bf 0412} (2004) 012
  [hep-lat/0408034].
%
\bibitem{Schleifenbaum:2004id}
  W.~Schleifenbaum, A.~Maas, J.~Wambach and R.~Alkofer,
  Phys.\ Rev.\ D {\bf 72} (2005) 014017
  [hep-ph/0411052].
 %%CITATION = PHRVA,D72,014017;%%

%\cite{Cucchieri:2008qm}
\bibitem{Cucchieri:2008qm}
  A.~Cucchieri, A.~Maas and T.~Mendes,
  %``Three-point vertices in Landau-gauge Yang-Mills theory,''
  Phys.\ Rev.\ D {\bf 77} (2008) 094510
  [arXiv:0803.1798 [hep-lat]].
  %%CITATION = ARXIV:0803.1798;%%
  
%\cite{Hopfer:2013np}
\bibitem{Hopfer:2013np}
  M.~Hopfer, A.~Windisch and R.~Alkofer,
  %``The Quark-Gluon Vertex in Landau gauge QCD,''
  PoS CONFINEMENTX {\bf } (2013) 073
  [arXiv:1301.3672 [hep-ph]].
  %%CITATION = ARXIV:1301.3672;%%  

%\cite{Alkofer:2008tt}
\bibitem{Alkofer:2008tt}
  R.~Alkofer, C.~S.~Fischer, F.~J.~Llanes-Estrada and K.~Schwenzer,
  %``The Quark-gluon vertex in Landau gauge QCD: Its role in dynamical chiral symmetry breaking and quark confinement,''
  Annals Phys.\  {\bf 324} (2009) 106
  [arXiv:0804.3042 [hep-ph]].
  %%CITATION = ARXIV:0804.3042;%%
  %90 citations counted in INSPIRE as of 28 Oct 2013

%\cite{Williams:2014iea}
\bibitem{Williams:2014iea}
  R.~Williams,
  %``The quark-gluon vertex in Landau gauge bound-state studies,''
  arXiv:1404.2545.
  %%CITATION = ARXIV:1404.2545;%%

%\cite{Aguilar:2014lha}
\bibitem{Aguilar:2014lha}
  A.~C.~Aguilar, D.~Binosi, D.~Iba\~{n}ez and J.~Papavassiliou,
  %``A new method for determining the quark-gluon vertex,''
  arXiv:1405.3506 [hep-ph].
  %%CITATION = ARXIV:1405.3506;%%

%\cite{Marciano:1977su}
\bibitem{Marciano:1977su}
  W.~J.~Marciano and H.~Pagels,
  %``Quantum Chromodynamics: A Review,''
  Phys.\ Rept.\  {\bf 36} (1978) 137.
  %%CITATION = PRPLC,36,137;%%
  %795 citations counted in INSPIRE as of 28 Oct 2013

%\cite{Ball:1980ay}
\bibitem{Ball:1980ay}
  J.~S.~Ball and T.~-W.~Chiu,
  %``Analytic Properties of the Vertex Function in Gauge Theories. 1.,''
  Phys.\ Rev.\ D {\bf 22} (1980) 2542;
  %%CITATION = PHRVA,D22,2542;%%
%\cite{Ball:1980ax}
% \bibitem{Ball:1980ax}
  J.~S.~Ball and T.~-W.~Chiu,
  %``Analytic Properties Of The Vertex Function In Gauge Theories. 2.,''
  Phys.\ Rev.\ D {\bf 22} (1980) 2550
   [Erratum-ibid.\ D {\bf 23} (1981) 3085].
  %%CITATION = PHRVA,D22,2550;%%


%\cite{Aguilar:2013ac}
\bibitem{Aguilar:2013ac}
  A.~C.~Aguilar, D.~Binosi, J.~C.~Cardona and J.~Papavassiliou,
  %``Nonperturbative results on the quark-gluon vertex,''
  PoS ConfinementX {\bf } (2012) 103
  [arXiv:1301.4057 [hep-ph]].

%\cite{Rojas:2013tza}
\bibitem{Rojas:2013tza}
  E.~Rojas, J.~P.~B.~C.~de Melo, B.~El-Bennich, O.~Oliveira and T.~Frederico,
  %``On the Quark-Gluon Vertex and Quark-Ghost Kernel: combining Lattice Simulations with Dyson-Schwinger equations,''
  JHEP {\bf 1310} (2013) 193
  [arXiv:1306.3022 [hep-ph]].
  %%CITATION = ARXIV:1306.3022;%%

%\cite{Taylor:1971ff}
\bibitem{Taylor:1971ff}
  J.~C.~Taylor,
  %``Ward Identities and Charge Renormalization of the Yang-Mills Field,''
  Nucl.\ Phys.\ B {\bf 33} (1971) 436.
  %%CITATION = NUPHA,B33,436;%%
  %624 citations counted in INSPIRE as of 28 Oct 2013

%\cite{vonSmekal:2009ae}
\bibitem{vonSmekal:2009ae}
  L.~von Smekal, K.~Maltman and A.~Sternbeck,
  %``The Strong coupling and its running to four loops in a minimal MOM scheme,''
  Phys.\ Lett.\ B {\bf 681} (2009) 336
  [arXiv:0903.1696 [hep-ph]].
  %%CITATION = ARXIV:0903.1696;%%
  %37 citations counted in INSPIRE as of 18 Sep 2014

%\cite{Lerche:2002ep}
\bibitem{Lerche:2002ep}
  C.~Lerche and L.~von Smekal,
  %``On the infrared exponent for gluon and ghost propagation in Landau gauge QCD,''
  Phys.\ Rev.\ D {\bf 65} (2002) 125006
  [hep-ph/0202194].
  %%CITATION = HEP-PH/0202194;%%
  %299 citations counted in INSPIRE as of 16 May 2014

%\cite{Pagels:1979hd}
\bibitem{Pagels:1979hd}
  H.~Pagels and S.~Stokar,
  %``The Pion Decay Constant, Electromagnetic Form-Factor and Quark Electromagnetic Selfenergy in QCD,''
  Phys.\ Rev.\ D {\bf 20} (1979) 2947.
  %%CITATION = PHRVA,D20,2947;%%
  
%\cite{Hopfer:2012ht}
\bibitem{Hopfer:2012ht}
  M.~Hopfer, R.~Alkofer and G.~Haase,
  %``Solving the Ghost-Gluon System of Yang-Mills Theory on GPUs,''
  Comput.\ Phys.\ Commun.\  {\bf 184} (2013) 1183
  [arXiv:1206.1779 [hep-ph]].
  %%CITATION = ARXIV:1206.1779;%%

%\cite{Fischer:2005en}
\bibitem{Fischer:2005en}
  C.~S.~Fischer, P.~Watson and W.~Cassing,
  %``Probing unquenching effects in the gluon polarisation in light mesons,''
  Phys.\ Rev.\ D {\bf 72} (2005) 094025
  [hep-ph/0509213].
  %%CITATION = HEP-PH/0509213;%%
  %40 citations counted in INSPIRE as of 22 Nov 2013

%\cite{Kizilersu:2009kg}
\bibitem{Kizilersu:2009kg}
  A.~Kizilersu and M.~R.~Pennington,
  %``Building the Full Fermion-Photon Vertex of QED by Imposing Multiplicative Renormalizability of the Schwinger-Dyson Equations for the Fermion and Photon Propagators,''
  Phys.\ Rev.\ D {\bf 79} (2009) 125020
  [arXiv:0904.3483 [hep-th]].
  %%CITATION = ARXIV:0904.3483;%%
  
%\cite{Curtis:1990zs}
\bibitem{Curtis:1990zs}
  D.~C.~Curtis and M.~R.~Pennington,
  %``Truncating the Schwinger-Dyson equations: How multiplicative renormalizability and the Ward identity restrict the three point vertex in QED,''
  Phys.\ Rev.\ D {\bf 42} (1990) 4165.
  %%CITATION = PHRVA,D42,4165;%%
  %205 citations counted in INSPIRE as of 30 Apr 2014

%\cite{Fischer:2002eq}
\bibitem{Fischer:2002eq}
  C.~S.~Fischer, R.~Alkofer and H.~Reinhardt,
  %``The Elusiveness of infrared critical exponents in Landau gauge Yang-Mills theories,''
  Phys.\ Rev.\ D {\bf 65} (2002) 094008
  [hep-ph/0202195].
  %%CITATION = HEP-PH/0202195;%%
  %134 citations counted in INSPIRE as of 27 Nov 2013
 
%\cite{Eichmann:2014xya}
\bibitem{Eichmann:2014xya}
  G.~Eichmann, R.~Williams, R.~Alkofer and M.~Vujinovic,
  %``The three-gluon vertex in Landau gauge,''
  Phys.\ Rev.\ D {\bf 89} (2014) 105014
  [arXiv:1402.1365 [hep-ph]];
  %%CITATION = ARXIV:1402.1365;%%
  %6 citations counted in INSPIRE as of 27 May 2014%\cite{Vujinovic:2014fza}
%\bibitem{Vujinovic:2014fza}
  M.~Vujinovic, R.~Alkofer, G.~Eichmann and R.~Williams,
  %``Non-perturbative features of the three-gluon vertex in Landau gauge,''
  arXiv:1404.4474 [hep-ph].
  %%CITATION = ARXIV:1404.4474;%%

%\cite{Blum:2014gna}
\bibitem{Blum:2014gna}
  A.~Blum, M.~Q.~Huber, M.~Mitter and L.~von Smekal,
  %``Gluonic three-point correlations in pure Landau gauge QCD,''
  Phys.\ Rev.\ D {\bf 89} (2014) 061703
  [arXiv:1401.0713 [hep-ph]].
  %%CITATION = ARXIV:1401.0713;%%
  %11 citations counted in INSPIRE as of 12 May 2014

%\cite{Huber:2014tva}
\bibitem{Huber:2014tva}
  M.~Q.~Huber and L.~von Smekal,
  %``Spurious divergences in Dyson-Schwinger equations,''
  arXiv:1404.3642 [hep-ph].
  %%CITATION = ARXIV:1404.3642;%%

%\cite{Press:1992zz}
\bibitem{press:nr2007}
  W.~H.~Press, S.~A.~Teukolsky, W.~T.~Vetterling and B.~P.~Flannery,
  ``Numerical Recipes: The Art of Scientific Computing,'' ISBN-10: 0521880688.


\end{thebibliography}
\end{document}